\documentclass{article}
\usepackage[a4paper, total={6.5in,9in}]{geometry}
\usepackage[utf8]{inputenc}
\usepackage{amsmath}
\usepackage{amssymb}
\usepackage{graphicx}
\usepackage{authblk}
\usepackage{xcolor}
\usepackage{url}
\usepackage[authoryear]{natbib}
\usepackage{algorithm2e}
\usepackage{appendix}
\usepackage{rotating}
\usepackage{appendix}
\usepackage{comment}
\usepackage[normalem]{ulem}
\usepackage{longtable}

\usepackage[english]{babel}
\usepackage[autostyle]{csquotes}
\usepackage{setspace}

\title{Uncertainty, volatility and the persistence norms of financial time series}
\vspace{-40pt}
\author[1]{Simon Rudkin\thanks{\textbf{Corresponding Author}. Full Address: Economics Department, School of Social Sciences, Swansea University, Bay Campus, Swansea, SA1 8EN, United Kingdom. Email:s.t.rudkin@swansea.ac.uk}}
\affil[1]{Economics Department, Swansea University, United Kingdom}
\vspace{-30pt}
\author[2]{Wanling Qiu \thanks{Full Address: Accounting and Finance Department, School of Management, Swansea University, Bay Campus, Swansea, SA1 8EN, United Kingdom. Email:wanling.qiu@swansea.ac.uk}}
\affil[2]{Accounting and Finance Department, Swansea University, United Kingdom}
\vspace{-30pt}
\author[1]{Pawe{\l} D{\l}otko\thanks{Full Address: Dioscuri Centre in Topological Data Anlaysis, Mathematics Institute, Polish Academy of Sciences, Warsaw, 01-2000, Poland. Email:pdlotko@impan.pl . P.D{\l}otko acknowledges support from 
the Dioscuri program initiated by the Max Planck Society, jointly managed by the National Science Centre (Poland), and mutually funded by the Polish Ministry of Science and Higher Education and the German Federal Ministry of Education and Research.}}
\affil[1]{Dioscuri Centre in Topological Data Analysis, Polish Academy of Sciences, Poland}

\vspace{-40pt}

\date{September 2021}

\begin{document}

\maketitle
\begin{abstract}
    Norms of Persistent Homology introduced in topological data analysis are seen as indicators of system instability, analogous to the changing predictability that is captured in financial market uncertainty indexes. This paper demonstrates norms from the financial markets are significant in explaining financial uncertainty, whilst macroeconomic uncertainty is only explainable by market volatility. Meanwhile, volatility is insignificant in the determination of norms when uncertainty enters the regression. Persistence norms therefore have potential as a further tool in asset pricing, and also as a means of capturing signals from financial time series beyond volatility.
\end{abstract}
\vspace{-10pt}

\section{Introduction}

Since \cite{gidea2018topological} identified the power of topological data analysis (TDA) persistence norms to warn about financial crashes, there has been strong interest in how the tools of TDA can be brought to finance and economics \citep{goel2020topological,bakas2020commodity,gidea2020topological,majumdar2020clustering}. TDA is introduced in the seminal work of \cite{carlsson2009topology}, and is widely developed for, the natural sciences \citep{bubenik2017persistence,bubenik2020persistence}. Initial potential for financial market dynamics of the type described by \cite{hsieh1991chaos} comes from the applications in the detection of phase transitions and system instability \citep{stolz2017persistent,smith2021topological}. Promising advancements on correlation and volatility show how each alter the topology of the time series, but that neither correlation or volatility alone can explain persistence norms \citep{aromi2021topological,leaverton2020some,katz2021time}. Focus in this literature has been on the strength of the volatility-norm relationship. We expand that focus to consider financial market dynamics, asking how persistence norms link to forward looking measures of uncertainty

\cite{jurado2015measuring} constructs uncertainty indexes for financial (FIN), macroeconomic (MAC) and real activity (REA) based upon the predictability of relevant series under each heading\footnote{We provide a discussion of the \cite{jurado2015measuring} uncertainty indexes within the supplementary material. The interested reader is directed to the \cite{jurado2015measuring} paper, and associated supplementary material, for full details on index construction.}. Of these, the MAC index sees most adoption in asset pricing \citep{bali2017economic,bali2021macroeconomic}. FIN has a natural association with the topological norms we derive from the financial markets, as our measure stems from the returns on four major US stock indices. This paper therefore focuses primarily on FIN and MAC. Regression of persistence norms on FIN yields an insignificant constant term, suggesting that FIN can span persistence norms. Further, adding volatility does not remove statistical significance from FIN. MAC is only significantly linked to volatility. Conversely, neither norms nor volatility span uncertainty\footnote{There are many combinations of uncertainty and model parameters that yield qualitatively similar results that are not reported in this paper for brevity. All results, including those for REA, for horizons of 3 months and 12 months, and for the economic elements of the uncertainty indexes, are available on request from the authors.}. 

This paper focuses on contemporaneous relationships between volatility, uncertainty and persistence norms in recognition of the way these series are used in the asset pricing literature \citep{ang2006cross,bali2017economic,bali2021macroeconomic}. Our empirical contribution is the demonstration that norms are better explained by FIN than volatility, but that norms and volatility alone do not directly explain uncertainty. In evaluating the links between \cite{jurado2015measuring} and persistence norms for the first time, this paper furthers the understanding of how TDA can enhance financial modelling\footnote{Code to allow the recreation of results in this paper is available from the website of Simon Rudkin at \texttt{https://sites.google.com/view/simonrudkin/tda-research}}.

\section{Persistence Norms}
Consider a point cloud $X = \{x_1,\ldots,x_n\}$ where each point $\{x_i\}_{i=1}^k$ have $k$ coordinates. Given $X$, an \emph{abstract simplicial complex} $\mathcal{K}$, being a family of subsets of $X$, will be constructed. Each subset consisting of points $x_{l_0},\ldots,x_{l_k}$ will constitute a $k-$dimensional simplex. Such a simplex is equipped with a \emph{filtration value} being equal to maximal distance between each pair of points among $\{x_{l_0},\ldots,x_{l_k}\}$. This distance will be referred to as a \emph{filtration of the simplex}.

Clearly, when each subsets of $X$ constitute a simplex, the size of the obtained data structure is $2^n$ and therefore not tractable computationally. Consequently, in what follows, only the simplices of a filtration value bounded by a predefined constant $r$ (and often of a bounded cardinality) will be considered in a restricted simplicial complex $\mathcal{K}_r$. Note that when a simplex constituted in points $\{x_{l_0},\ldots,x_{l_k}\}$ belongs to $\mathcal{K}_r$, each simplex supported in a subset of $\{x_{l_0},\ldots,x_{l_k}\}$ will also belong to $\mathcal{K}_r$.

An obtained simplicial complex $\mathcal{K}_r$ is called a \emph{Rips complex} of $X$. Let us consider the simplices of $\mathcal{K}_r$ in order of their filtration. Concentrating on connected components, we can observe creation of new connected components, and merging of existing ones as the simplices of higher and higher filtration values are considered. The filtration value on which the components connect will be called \emph{birth}, while the filtration value at which the connected components connect to another component is the \emph{death} of the original connected component. However, the multi-scale approach presented here is not restricted to connected components. When considering higher dimensional topological features like holes and voids, analogously each feature can be equipped with a birth and a death filtration value. Note that some of the features will not die in $\mathcal{K}_r$, in which case, they are said to have an infinite length. This assignment is a basis of \emph{persistent homology}. For more formal, yet intuitive introduction to the subject please consider~\cite{carlsson2009topology}

The life-time of a feature is calculated as the death minus the birth value. Note that in what follows, we only consider features of a finite lifetime. Life information can be combined into $L_1$ being the sum of the lives of the features, whilst the $L_2$ norm is the square root of the sum of the squared lives of the features. That is:
\begin{align}
    L_{11} = \sum_{f=1}^{F_0} d_f-b_f \\
    L_{12} = \sqrt{\sum_{f=1}^{F_1} (d_f-b_f)^2 }
\end{align}
where feature $f$ is born at radius $b_f$ and dies at radius $d_f$. For dimension as $z$, $z=0,1$, $F_z$ is the total number of features identified. Note that so called 1-persistent landscape norms used in~\cite{gidea2018topological} are simply $L_{11}$ up to a constant re-scaling, hence we will focus on them in the further work. 

Influences on the norm are the distances between points within the cloud and the number of points in the cloud. For comparability all of the clouds in this paper contain the similar numbers of points meaning any difference may be attributed to the distance between points. Distances are impacted by the correlation between axes and the variance within the underlying distributions that generate the cloud. Volatility and the correlation between market index returns capture those distributional features in this paper.

\section{Data}

Monthly uncertainty data is obtained from the website of Sydney Ludvigsson\footnote{Downloaded 12th September 2021 from \texttt{https://www.sydneyludvigson.com/data-and-appendixes}.}. Daily index data is downloaded through Yahoo! Finance with daily log returns for day $t$, $R_t$, constructed from adjusted closing prices, $p_t$, using $R_t = \ln{p_t} - \ln{p_{t-1}}$. Monthly returns may then be aggregated across all days in the month. Using the daily data we construct point clouds of length 1 month, 3 months, 6 months and 12 months and compute the persistence norms there upon. Following \cite{gidea2018topological} a cloud is produced using log returns on the S\&P 500, Dow Jones Industrial Average, NASDAQ and Russell 2000. The geometric mean of the standard deviation of returns on the four indices provides the measure of volatility\footnote{The shape of the point cloud is dependent on the variance-covariance matrix of the input series but is not order dependent. Further each index is weighted equally in the construction of the persistence norms. Therefore the average volatility is seen as a parsimonious measure to apply in this paper. See \cite{aromi2021topological} for a discussion of the role of the variance-covariance matrix and empirical examples using the four index cloud considered here.}. Allowing for availability of stock index returns and uncertainty measures, our sample runs from 1st January 1993 to 28th June 2021.

\begin{table}
	\begin{center}
		\caption{Correlations between index return volatility, persistence norms and uncertainty}
		\label{tab:corunc}
		\begin{tabular}{l l c c c c c c c c c  c c c c}
			\hline
			&& $FIN1$ & $MAC1$ & $REA1$ & $\bar{\sigma}$ & $L_{11}$ & $L_{12}$ && $FIN1$ & $MAC1$ & $REA1$ & $\bar{\sigma}$ & $L_{11}$ & $L_{12}$\\
			\hline
			&&&&&&&&&&\\
			& \multicolumn{6}{l}{Panel (a) 1 Month Cloud:} && \multicolumn{6}{l}{Panel (b) 3 Month Cloud:}\\ &$\bar{\sigma}$ & 0.743&0.585&0.413&1&&&&0.804&0.683&0.515&1&&\\
			& $L_{11}$ & 0.147&0.100&0.167&0.059&1&&&0.532&0.287&0.247&0.465&1&\\
			&$L_{12}$&0.185&0.117&0.174&0.113&0.967&1&&0.602&0.379&0.319&0.586&0.940&1\\
			&&&&&\\
			&& \multicolumn{6}{l}{Panel (c) 6 Month Cloud:} && \multicolumn{6}{l}{Panel (d) 12 Month Cloud:}\\
			&$\bar{\sigma}$ &0.814&0.721&0.560&1&&&&0.789&0.697&0.547&1&&\\
			& $L_{11}$ & 0.592&0.304&0.240&0.551&1&&&0.597&0.288&0.188&0.617&1&\\
			& $L_{12}$&0.663&0.403&0.318&0.653&0.964&1&&0.665&0.381&0.264&0.715&0.982&1\\
			\hline
		\end{tabular}
	\end{center}
	\footnotesize{Notes: Figures report correlations between stated row variables and columns. $FIN1$, $MAC1$ and $REA1$ are the 1 month ahead financial, macroeconomic and real uncertainty index measures from \cite{jurado2015measuring} as downloaded from the website of Sydney Ludvigsson at \texttt{https://www.sydneyludvigson.com/data-and-appendixes}. Point clouds are constructed from the daily returns on four stock indices, the S\&P 500, NASDAQ, the Dow Jones Industrial Average and the Russell 2000. Cloud lengths are indicated as the panel names. $\bar{\sigma}$ is the average volatility from the four stock indices. $L_{11}$ and $L_{12}$ are the dimension 1 $L_1$ and $L_2$ persistence norms respectively. Pairwise correlations within the \cite{jurado2015measuring} uncertainty index rows are omitted as they are not affected by cloud length. Sample from 1st January 1993 to 28th June 2021.}  
\end{table}

Correlations between the \cite{jurado2015measuring} uncertainty indexes and volatility are high, being approximately 0.7 for $FIN1$, around 0.6 for $MAC1$ and approximately 0.6 for $REAL$. Correlations between volatility and the uncertainty indexes are always higher than the corresponding correlations between uncertainty and the persistence norms. Table \ref{tab:corunc} shows correlations between $L_1$ norms and $FIN1$ to be below 0.6 in the 3-month and 6-month clouds, rising to near 1 for the 12-month cloud\footnote{We present a discussion of the correlation between measures of volatility and the persistence norms across the cloud lengths in the supplementary material.}. $L_2$ norms have a correlation of just above 0.6 with $FIN1$. With $MAC1$ we see correlations below 0.4, and with $REA1$ the correlations are lower at 0.3 or less. Whilst the evidence suggests uncertainty is picking up much of the volatility in the cloud, the norms are providing different information about the data. A note of caution is raised by the strong correlation between volatility and $FIN1$ when using these two metrics to model persistence norms. When $MAC1$ is used the correlations with volatility are lower meaning fewer modelling concerns.

\begin{figure}
    \begin{center}
        \caption{Persistence norms, volatility and uncertainty time series plots}
        \label{fig:unc1}
        \begin{tabular}{c c}
            \includegraphics[width=8cm]{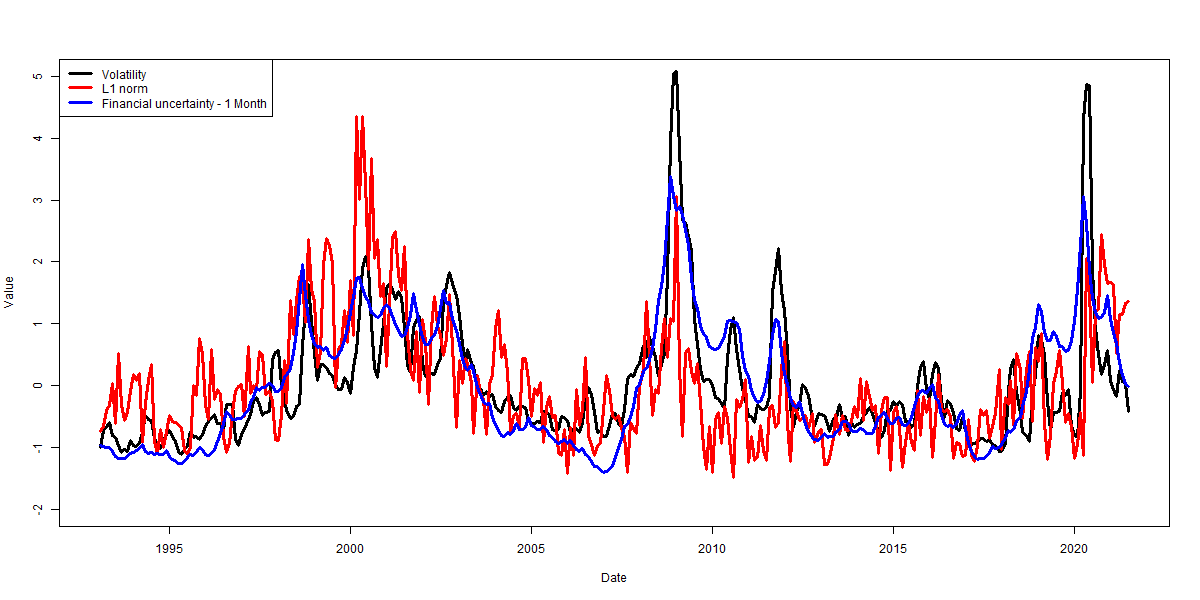} & \includegraphics[width=8cm]{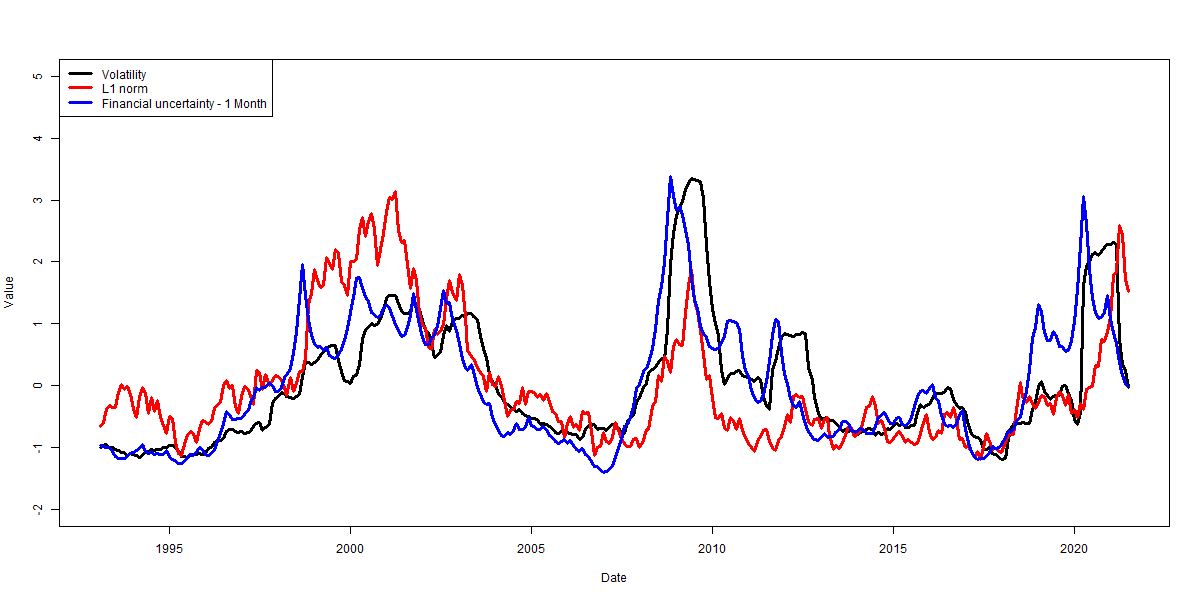}  \\
            (a) $FIN1$  3 months & (b) $FIN1$ 12 months  \\ \includegraphics[width=8cm]{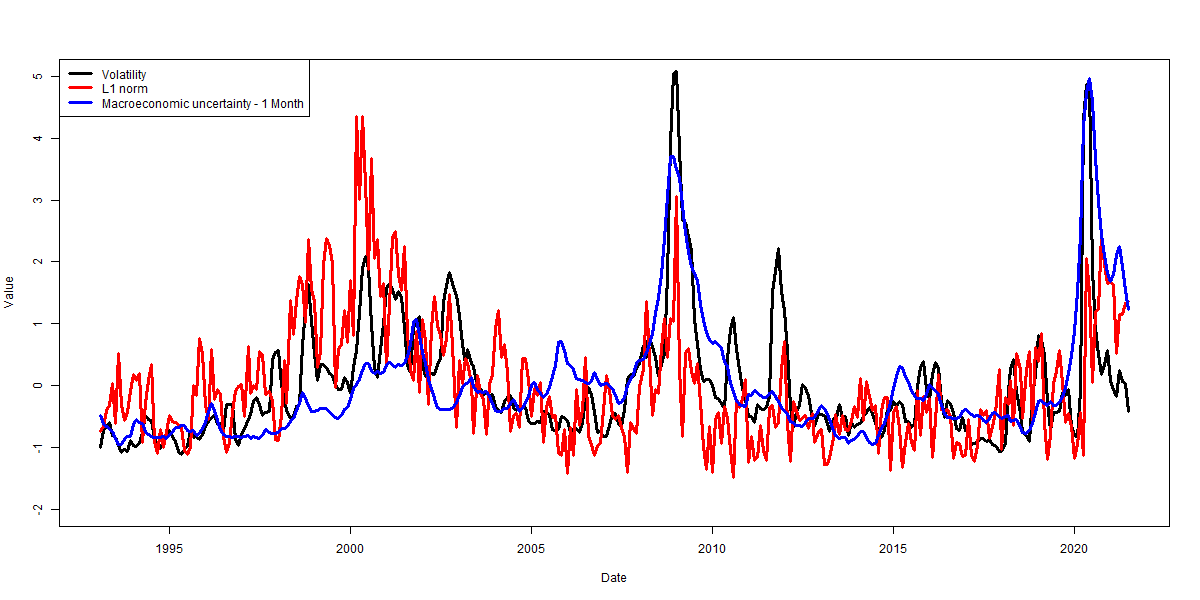} & \includegraphics[width=8cm]{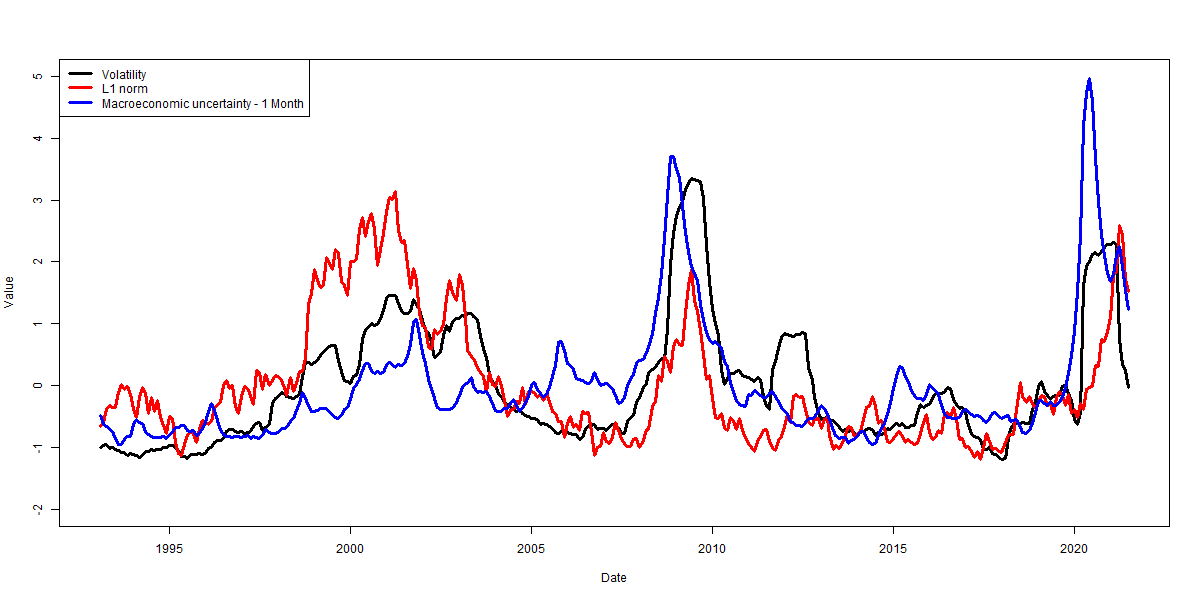}  \\
            (a) $MAC1$ 3 months & (b) $MAC1$ 12 months  \\
        \end{tabular}
    \end{center}
\footnotesize{Notes: Figures plot the financial (FIN) and macroeconomic (MAC) uncertainty index of \cite{jurado2015measuring}. Each plot uses the 1 month ahead uncertainty index $FIN1$ and $MAC1$. These indices are plotted as blue lines. On each plot the average return volatility of the four indexes within the cloud; being the S\&P500, Dow Jones Industrial Average, NASDAQ and the Russell 2000 index. Average volatility is plotted in black. The dimension 1 $L_2$ persistence norm from these clouds are shown in red. Values are standardised by subtracting the mean value and dividing by the standard deviation. Size of window used in the estimation is provided below each plot. Sample from 1st January 1993 to 28th June 2021.}
\end{figure}

Plotting the uncertainty, volatility and $L_1$ persistence norms series, Figure \ref{fig:unc1} shows that whilst using more lags, or considering longer time periods produces more consistency, these are still volatile series. Choice of $L_1$ follows \cite{gidea2018topological}, but the plots with $L_2$ included in the supplementary material are very similar as the high correlation between $L_1$ and $L_2$ would suggest. Across the full sample it is difficult to conclude which series can be used to explain the others\footnote{Developing lead-lag relationships within this data is left to future work, with greater focus needed on TDA measures from the same series as the uncertainty measures.}. Panels (a) and (b) show the persistence norms being high during the ``dot-com'' bubble bursting in the early part of this century, a time where uncertainty is high but volatility is not. Meanwhile for both the Covid-19 pandemic and the global financial crisis we see volatility and uncertainty go to high standardised values whilst norms remain comparatively low. Panels (c) and (d) show closer alignment between $MAC1$ and volatility since $MAC1$ does not peak in the ``dot-com'' bubble like $FIN1$. These observations are important for evaluating the use of uncertainty, and particularly $MAC1$ in the asset pricing literature \citep{bali2017economic,bali2021macroeconomic}. Consequently the time series plots show potential for $L_1$ norms in financial uncertainty. 

\section{Results}

Our results are presented for the dimension one $L_1$ norm to keep this work in line with \cite{gidea2018topological}. Results for the corresponding dimension 1 $L_2$ norms are provided in the supplementary material.

\subsection{Persistence Norms}

Regression of persistence norms on the uncertainty indices of \cite{jurado2015measuring} and the average volatility index returns tests whether these financial market variables are able to span the persistence norms. Formally we estimate a total of 216 models which are based upon three base regressions\footnote{For each measure of uncertainty there are 4 point clouds to consider. Three uncertainty indices at three different lags, and with both a standard and economic version for each of the 9, this all means 72 combinations arise from Model 1. Note that there can only be one of the 18 possible uncertainty indexes and one of the four cloud lengths used in any one regression. Likewise we can only use one of the two norms, either $L_1$ or $L_2$ in a regression. Each model therefore comprises one uncertainty index, one norm and/or one volatility.}: 
\begin{align}
    \mathtt{Model 1:} & L_{11} = \alpha_1 + \beta_{11}FIN1 + \epsilon \\
    \mathtt{Model 2:} & L_{11} = \alpha_2 + \beta_{21}\bar{\sigma} + \epsilon \\
    \mathtt{Model 3:} & L_{11} = \alpha_3 + \beta_{31}FIN1 + \beta_{32}\sigma + \epsilon 
\end{align}
Where $L_{11}$ is the dimension one $L_1$ persistence norm, $\sigma$ is the standard deviation of S\&P 500 returns and $FIN1$ is the one month ahead financial uncertainty index of \cite{jurado2015measuring}. In all cases $\epsilon$ is the error term. In all estimations \cite{newey1987simple} adjusted standard errors are employed to account for autocorrelation and heteroskedasticity inherent in financial time series. 

\begin{sidewaystable}
    \begin{center}
        \caption{Determinants of $L_1$ persistence norms}
        \label{tab:reg41}
        \begin{tabular}{ll l c c c c c c c c c c c c c}
        \hline
              &&& \multicolumn{4}{l}{Model 1} & \multicolumn{4}{l}{Model 2}& \multicolumn{4}{l}{Model 3}\\
             &&& 1m & 3m & 6m & 12m & 1m & 3m & 6m & 12m & 1m & 3m & 6m & 12m\\
             \hline
             \multicolumn{12}{l}{Panel (a): Financial Uncertainty ($FIN1$): }\\
             & Constant & $\alpha$ & 0.031&-0.308&-0.530&-0.354&0.133&0.511&1.171&2.365&-0.005&-0.206&-0.151&0.727\\
              &&&(0.574)&(1.270)&(0.674)&(0.150)&(7.680)&(6.103)&(3.780)&(2.525)&(0.083)&(0.659)&(0.225)&(0.484)\\
             & Uncertainty & $FIN1$ & 0.132&1.385&3.156&5.992&&&&&0.206&1.163&2.262&2.926\\
             &&&(2.012)&(4.504)&(3.091)&(1.830)&&&&&(2.501)&(2.453)&(2.162)&(1.243)\\
             & Volatility & $\bar{\sigma}$&&&&&0.015&0.378&0.991&2.281&-0.029&0.086&0.370&1.431\\
             &&&&&&&(0.952)&(4.275)&(2.961)&(2.069)&(1.351)&(0.885)&(1.558)&(2.756)\\
             & \multicolumn{2}{l}{Adjusted $R^2$}&0.022&0.283&0.350&0.356&0.004&0.216&0.304&0.381&0.027&0.287&0.364&0.413\\
             &&&&&&&\\
             \multicolumn{9}{l}{Panel (b): Macroeconomic Uncertainty ($MAC1$): }\\
             & Constant & $\alpha$ & 0.058&0.174&0.654&2.083&0.133&0.511&1.171&2.365&0.058&0.629&1.948&4.378\\
            &&&(0.702)&(0.907)&(0.801)&(0.684)&(7.680)&(6.103)&(3.780)&(2.525)&(0.608)&(1.527)&(1.516)&(2.616)\\
            & Uncertainty & $MAC1$ & 0.143&1.190&2.580&4.601&&&&&0.141&-0.238&-1.660&-4.424\\
            &&&(1.053)&(3.969)&(1.958)&(0.912)&&&&&(0.764)&(0.271)&(0.553)&(1.007)\\
            & Volatility & $\bar{\sigma}$&&&&&0.015&0.378&0.991&2.281&0.000&0.410&1.244&2.995\\
            &&&&&&&(0.952)&(4.275)&(2.961)&(2.069)&(0.017)&(2.512)&(1.907)&(1.868)\\
            & \multicolumn{2}{l}{Adjusted $R^2$} & 0.010&0.082&0.092&0.083&0.004&0.216&0.304&0.381&0.010&0.218&0.322&0.420\\ 
             \hline
        \end{tabular}
    \end{center}
\footnotesize{Notes: Regressions of the stated outcome with the other two variables as explanatory factors. 3m and 6m refers to the number of months of data in the point cloud over which the persistence norms and volatility are calculated. Point clouds are the daily returns on the S\&P 500, Dow Jones Industrial Average, NASDAQ and Russell 2000 indexes. Uncertainty indexes are the 1 month ahead estimates from \cite{jurado2015measuring} as downloaded from the website of Sydney Ludvigson. Figures are the estimated coefficients and figures in parentheses are the associated \cite{newey1987simple} adjusted t-statistics. Model 1 has $L_{11} = \alpha_1 + \beta_{11}FIN1 + \epsilon$. Model 1 has $L_{11} = \alpha_2 + \beta_{21}\bar{\sigma} + \epsilon$. Model 3 has $L_{2} = \alpha_3 + \beta_{31}FIN1 + \beta_{32}\bar{\sigma} + \epsilon$. Models 1, 2 and 3 represent the base specification upon which all regression models are built. Specifications show $FIN1$ as the one month ahead financial uncertainty (FIN) index. Where macroeconomic uncertainty (MAC) is used then the equations for the four models are duly updated to replace $FIN1$ with $MAC1$. In each case $\bar{\sigma}$ is the average standard deviation of the four index returns in the window and $\epsilon$ is the error term. Estimation employs \cite{newey1987simple} robust standard errors to control for autocorrelation and heteroskedasticity inherent in these time series. Sample from 1st January 1993 to 28th June 2021. $n=342$}
\end{sidewaystable}

An immediate inference from Table \ref{tab:reg41} is that regressions involving the uncertainty indexes do not have significant constant terms. Further, coefficients on FIN1 are significant in all but model 3 on the 12-month cloud. Meanwhile, volatility is not significant when uncertainty is added. The significant constants in Model 2 confirm that the norm of the four index cloud is picking up additional information beyond the volatility in the cloud. When MAC1 is used as the uncertainty measure, there is some significance in model 1 for the 3-month and 6-month clouds, but model 3 has no significant coefficients.

\subsection{Uncertainty}

Reversing the question we regress the uncertainty indexes on the persistence norms and volatility. 
\begin{align}
    \mathtt{Model 4:} & FIN1 = \alpha_4 + \beta_{41}L_{11} + \epsilon \\
    \mathtt{Model 5:} & FIN1 = \alpha_5 + \beta_{52}\bar{\sigma} + \epsilon\\
    \mathtt{Model 6:} & FIN1 = \alpha_6 + \beta_{61}L_{11} + \beta_{62}\bar{\sigma} + \epsilon
\end{align}
Notation follows models 1 to 3.

\begin{sidewaystable}
    \begin{center}
        \caption{Explaining the uncertainty index}
        \label{tab:reg41}
        \begin{tabular}{ll l c c c c c c c c c c c c c}
        \hline
             &&& \multicolumn{4}{l}{Model 4} & \multicolumn{4}{l}{Model 5}& \multicolumn{4}{l}{Model 6}\\
             &&& 1m & 3m & 6m & 12m & 1m & 3m & 6m & 12m & 1m & 3m & 6m & 12m\\
             \hline
             \multicolumn{12}{l}{Panel (a): Financial Uncertainty ($FIN1$): }\\
             & Constant & $\alpha$ & 0.877&0.710&0.645&0.602&0.667&0.617&0.584&0.560&0.651&0.577&0.539&0.518\\
            &&&(10.76)&(17.72)&(11.39)&(6.192)&(20.77)&(16.46)&(15.70)&(8.128)&(20.94)&(16.51)&(14.95)&(6.359)\\
            & Norms & $L_{11}$&0.163&0.204&0.111&0.059&&&&&0.115&0.078&0.038&0.018\\
            &&& (2.422)&(8.155)&(6.500)&(3.641)&&&&&(2.445)&(3.431)&(2.149)&(1.001)\\
            & Volatility & $\bar{\sigma}$&&&&&0.213&0.251&0.274&0.290&0.211&0.222&0.236&0.250\\
            &&&&&&&(11.02)&(10.50)&(7.333)&(3.727)&(10.91)&(10.80)&(6.070)&(2.800)\\
            & \multicolumn{2}{l}{Adjusted $R^2$}&0.022&0.283&0.350&0.356&0.552&0.646&0.663&0.622&0.563&0.678&0.693&0.642\\
             &&&&&&&\\
             \multicolumn{9}{l}{Panel (b): Macroeconomic Uncertainty ($MAC1$): }\\
             & Constant & $\alpha$ & 0.634&0.580&0.562&0.554&0.528&0.492&0.468&0.455&0.522&0.497&0.487&0.489\\
            &&&(8.488)&(25.77)&(17.88)&(10.97)&(31.34)&(14.50)&(9.186)&(7.774)&(29.43)&(13.01)&(9.998)&(9.931)\\
            & Norms & $L_{11}$ & 0.070&0.069&0.036&0.018&&&&&0.046&-0.009&-0.016&-0.014\\
            &&&(0.618)&(1.041)&(1.260)&(1.096)&&&&&(0.725)&(0.274)&(0.649)&(0.981)\\
            & Volatility & $\bar{\sigma}$&&&&&0.105&0.134&0.153&0.161&0.105&0.138&0.168&0.194\\
            &&&&&&&(5.057)&(2.614)&(2.311)&(2.268)&(4.908)&(3.274)&(2.798)&(1.932)\\
            & \multicolumn{2}{l}{Adjusted $R^2$}&0.010&0.082&0.092&0.083&0.342&0.467&0.520&0.486&0.346&0.468&0.533&0.519\\
             \hline
        \end{tabular}
    \end{center}
\footnotesize{Notes: Regressions of the stated outcome with the other two variables as explanatory factors. 1m, 3m, 6m and 12m refers to the number of months of data in the point cloud over which the persistence norms and volatility are calculated. Uncertainty indexes are the 1 month ahead estimates from \cite{jurado2015measuring} as downloaded from the website of Sydney Ludvigson. Clouds are formed from the daily log returns on the S\&P 500, Dow Jones Industrial Average, NASDAQ and Russell 2000 stock market indexes.. Figures are the estimated coefficients and figures in parentheses are the associated \cite{newey1987simple} adjusted t-statistics. Model 4 has $FIN1 = \alpha_4 + \beta_{41}L_{12} + \epsilon$. Model 5 has $FIN1 = \alpha_5 + \beta_{51}\bar{\sigma} + \epsilon$. Model 6 has $FIN1 = \alpha_6 + \beta_{61}L_{12} + \beta_{62}\bar{\sigma}+\epsilon$. Specifications show $FIN1$ as the one month ahead financial uncertainty (FIN) index. Where macroeconomic uncertainty (MAC) is used then the equations for the four models are duly updated to replace $FIN1$ with $MAC1$. In each case $\bar{\sigma}$ is the average standard deviation of the returns on the four indices in the window. $\epsilon$ is the error term. Estimation employs \cite{newey1987simple} robust standard errors to control for autocorrelation and heteroskedasticity inherent in these time series. Sample from 1st January 1993 to 28th June 2021. $n=342$}
\end{sidewaystable}

Table \ref{tab:reg41} demonstrates the significance of volatility to the uncertainty index. $FIN1$ has significant terms on the $L_{11}$ norms in both Model 4 and Model 6. Persistence norms are also picking up an element of uncertainty in the market over and above that induced by volatility. For $MAC1$ any fit benefit of adding $L_1$ norms, model 6, to the univariate model with just volatility, model 5, is marginal. In all cases the constants are significant meaning that there are other features in the uncertainty index that the point cloud measures do not pick up.

\subsection{High and Low Uncertainty}

A feature of the uncertainty indexes of \cite{jurado2015measuring} is that they have very clear peaks. In the case of $FIN1$, Figure \ref{fig:unc1} showed that there were three peaks within our sample. $MAC1$ has peaks in the global financial crisis and now in the Covid-19 period. Therefore much of the variation in these indexes can be explained by factors that are effective at explaining the peaks. To explore this further we split the sample into those months where each uncertainty index is above, or below, its respective median. The 342 months of our sample therefore move into two sets of 171. 

\begin{sidewaystable}
    \begin{center}
        \caption{High and low uncertainty index periods}
        \label{tab:reg4hl}
        \begin{small}
        \begin{tabular}{ll l c c c c c c c c c c c c c}
        \hline
             &&& \multicolumn{4}{l}{Model 4} & \multicolumn{4}{l}{Model 5}& \multicolumn{4}{l}{Model 6}\\
             &&& 1m & 3m & 6m & 12m & 1m & 3m & 6m & 12m & 1m & 3m & 6m & 12m\\
             \hline
             \multicolumn{12}{l}{Panel (a): High Financial Uncertainty ($FIN1$): }\\
             & Constant & $\alpha$ & 1.054&0.963&0.943&0.945&0.877&0.836&0.821&0.822&0.867&0.808&0.792&0.806\\
             &&&(10.90)&(16.51)&(12.70)&(8.781)&(25.29)&(20.84)&(18.08)&(9.144)&(24.97)&(15.53)&(13.75)&(7.132)\\
             & Norms & $L_{11}$&0.044&0.084&0.042&0.019&&&&&0.056&0.034&0.016&0.005\\
             &&&(0.516)&(3.754)&(2.633)&(1.176)&&&&&(1.195)&(1.234)&(0.879)&(0.283)\\
             & Volatility & $\bar{\sigma}$&&&&&0.128&0.151&0.158&0.157&0.128&0.143&0.148&0.149\\
             &&&&&&&(7.316)&(7.887)&(4.930)&(2.144)&(7.230)&(7.106)&(4.141)&(1.740)\\
             & \multicolumn{2}{l}{Adjusted $R^2$} & 0.004&0.116&0.114&0.086&0.493&0.528&0.442&0.313&0.500&0.545&0.457&0.317\\
             &&&&&&&\\
             \multicolumn{12}{l}{Panel (b): Low Financial Uncertainty ($FIN1$): }\\
             & Constant & $\alpha$ & 0.741&0.739&0.694&0.643&0.678&0.625&0.579&0.62&0.678&0.627&0.564&0.575\\
             &&&(17.05)&(14.28)&(12.44)&(7.811)&(28.10)&(18.77)&(25.68)&(18.85)&(27.95)&(19.42)&(18.90)&(7.302)\\
             & Norms & $L_{11}$&0.007&0.003&0.027&0.025&&&&&-0.004&-0.003&0.011&0.013\\
             &&&(0.212)&(0.130)&(1.142)&(1.261)&&&&&(0.119)&(0.173)&(0.600)&(0.651)\\
             & Volatility & $\bar{\sigma}$&&&&&0.084&0.151&0.207&0.148&0.084&0.151&0.201&0.137\\
             &&&&&&&(3.626)&(4.062)&(8.092)&(4.490)&(3.631)&(4.457)&(7.064)&(3.648)\\
             &\multicolumn{2}{l}{Adjusted $R^2$}&0.000&0.000&0.046&0.092&0.142&0.257&0.392&0.348&0.142&0.257&0.399&0.374\\
             &&&&&&&&\\
             \multicolumn{9}{l}{Panel (c): High Macroeconomic Uncertainty ($MAC1$): }\\
             & Constant & $\alpha$ & 0.703&0.657&0.647&0.648&0.598&0.549&0.515&0.507&0.582&0.552&0.537&0.542\\
             &&&(8.618)&(23.02)&(16.26)&(10.47)&(24.77)&(16.95)&(11.38)&(7.552)&(28.05)&(14.31)&(13.17)&(9.276)\\
             & Norms & $L_{11}$ & 0.098&0.060&0.028&0.013&&&&&0.103&-0.006&-0.018&-0.016\\
             &&&(1.191)&(0.77)&(0.856)&(0.746)&&&&&(1.935)&(0.119)&(0.652)&(0.891)\\
             & Volatility & $\bar{\sigma}$&&&&&0.089&0.122&0.146&0.149&0.090&0.124&0.163&0.184\\
             &&&&&&&(4.158)&(3.217)&(2.921)&(2.042)&(3.991)&(3.670)&(3.362)&(1.688)\\
             & \multicolumn{2}{l}{Adjusted $R^2$}&0.021&0.069&0.062&0.045&0.281&0.428&0.504&0.407&0.304&0.428&0.523&0.450\\
             &&&&&&&\\
             \multicolumn{9}{l}{Panel (d): Low Macroeconomic Uncertainty ($MAC1$): }\\
             & Constant & $\alpha$ & 0.569&0.557&0.554&0.556&0.553&0.547&0.543&0.537&0.553&0.545&0.542&0.541\\
             &&&(54.13)&(39.02)&(26.33)&(28.50)&(57.69)&(40.03)&(40.67)&(30.85)&(57.10)&(34.82)&(31.26)&(32.77)\\
             & Norms & $L_{11}$ & 0.019&0.017&0.008&0.003&&&&&0.002&0.005&0.001&-0.002\\
             &&&(1.199)&(2.176)&(1.465)&(1.291)&&&&&(0.189)&(0.653)&(0.148)&(0.453)\\
             & Volatility & $\bar{\sigma}$ &&&&&0.025&0.031&0.031&0.034&0.025&0.028&0.028&0.033\\
             &&&&&&&(3.787)&(3.249)&(3.57)&(2.116)&(3.548)&(2.867)&(2.879)&(1.350)\\
             & \multicolumn{2}{l}{Adjusted $R^2$}&0.015&0.079&0.078&0.048&0.135&0.191&0.222&0.260&0.136&0.196&0.223&0.272\\
             \hline
        \end{tabular}
        \end{small}
    \end{center}
\footnotesize{Notes: Regressions of the stated outcome with the other two variables as explanatory factors. 1m, 3m, 6m and 12m refers to the number of months of data in the point cloud over which the persistence norms and volatility are calculated. Panels (a) and (c) cover periods of high financial and macroeconomic uncertainty respectively. Likewise panels (b) and (d) cover periods where uncertainty is below the sample median. Uncertainty indexes are the 1 month ahead estimates from \cite{jurado2015measuring} as downloaded from the website of Sydney Ludvigson. Clouds are formed from the daily log returns on the S\&P 500, Dow Jones Industrial Average, NASDAQ and Russell 2000 stock market indexes. Figures are the estimated coefficients and figures in parentheses are the associated \cite{newey1987simple} adjusted t-statistics. Model 4 has $FIN1 = \alpha_4 + \beta_{41}L_{11} + \epsilon$. Model 5 has $FIN1 = \alpha_5 + \beta_{51}\bar{\sigma} + \epsilon$. Model 6 has $FIN1 = \alpha_6 + \beta_{61}L_{11} + \beta_{62}\bar{\sigma}+\epsilon$. Specifications show $FIN1$ as the one month ahead financial uncertainty (FIN) index. Where macroeconomic uncertainty (MAC) is used then the equations for the four models are duly updated to replace $FIN1$ with $MAC1$. In each case $\bar{\sigma}$ is the average standard deviation of the returns on the four indices in the window. $\epsilon$ is the error term. Estimation employs \cite{newey1987simple} robust standard errors to control for autocorrelation and heteroskedasticity inherent in these time series. The number of observations in each high and low period is 171. Overall sample from 1st January 1993 to 28th June 2021.}
\end{sidewaystable}
   
Results in Table \ref{tab:reg4hl} derive from the estimation of models 4 to 6 on high, and low, uncertainty subsamples. In all cases the constant is highly significant for both FIN and MAC. Persistence norms are significant in high $FIN1$ periods when the 3-month and 6-month clouds, but this disappears after the addition of volatility into the model. Volatility is significant in the regressions of both $FIN1$ and $MAC1$, the significance robust to the addition of persistence norms to the model. Hence, whilst there is evidence of norms adding power in explaining uncertainty for the full sample, the split at the median shows the power to be insufficient in either high or low uncertainty periods.

\section{Summary}

Taking financial markets as complex dynamical systems \citep{hsieh1991chaos}, persistence norms provide a means to evaluate phase transitions and forewarn of crashes \citep{gidea2018topological}. An intuitive link to uncertainty in the spirit of \cite{jurado2015measuring} is made in this paper. Through contemporaneous regressions amongst the norms, volatility and uncertainty it has been shown that FIN, not volatility, has the strongest association with persistence norms. That persistence norms have explanatory power for FIN, but cannot span uncertainty, positions norms as a potentially useful further metric to add to the asset pricing arsenal. Although links between MAC and volatility shown in our results lend interpretation to the observed mispricing of MAC in \cite{bali2017economic}, there is only limited significance for persistence norms. Further development of the norms and volatility from the same dataset as the uncertainty indexes is a natural next step. Given the demonstrated roles of uncertainty and volatility in asset pricing, and the fact that volatility cannot explain the persistence norms, there is strong potential to also explore TDA signals in asset pricing.

\bibliography{unc}
\bibliographystyle{apalike}

\appendix

\setcounter{figure}{0}
\renewcommand{\thefigure}{A\arabic{figure}}
\setcounter{table}{0}
\renewcommand{\thetable}{A\arabic{table}}

\section{Measures of Uncertainty}

\cite{jurado2015measuring} provide three categories of uncertainty index, financial (FIN), macroeconomic (MAC) and real (REA), each based upon the predictability of a different set of time series. On an individual series uncertainty may be understood as the forecast error at a $h$-month horizon given information available at time $t$. Recognising that the information set $I_t$ available to investors is far wider than a single series it is necessary to use a large set of series to fully measure uncertainty. \cite{jurado2015measuring}'s approach seeks a common component of uncertainty to model from the observed data, rather than being based upon gathering large volumes of internet or newspaper data like the economic policy uncertainty index of \cite{baker2016measuring}. From the perspective of the main paper, the uncertainty indexes are taken as givens but it remains helpful to recognise the debate that informs their construction.

Details of the series and processes can be accessed via the paper and via the website of Sydney Ludvigson. In most asset pricing applications it is the MAC that is used, with the one-month ahead index (MAC1) being the choice for \cite{bali2017economic,bali2021macroeconomic}. However all three sets include a one-month, three-month and twelve-month ahead index. There are also versions of these indices computed for the subset of series which excludes any health related outcomes. This is referred to in the data descriptions as ``economic''. In the main paper we focus on MAC1 in line with the asset pricing literature, but also use FIN1 as the indicator derived from the financial markets directly. Here we demonstrate the tight relationships between the indexes, expanding the demonstration in \cite{bali2017economic} and the exposition in \cite{lee2021uncertainty}. 

\begin{figure}
    \begin{center}
        \caption{Comparison of Uncertainty Measures}
        \label{fig:aunc}
        \begin{tabular}{c c}
             \multicolumn{2}{c}{\includegraphics[width=12cm]{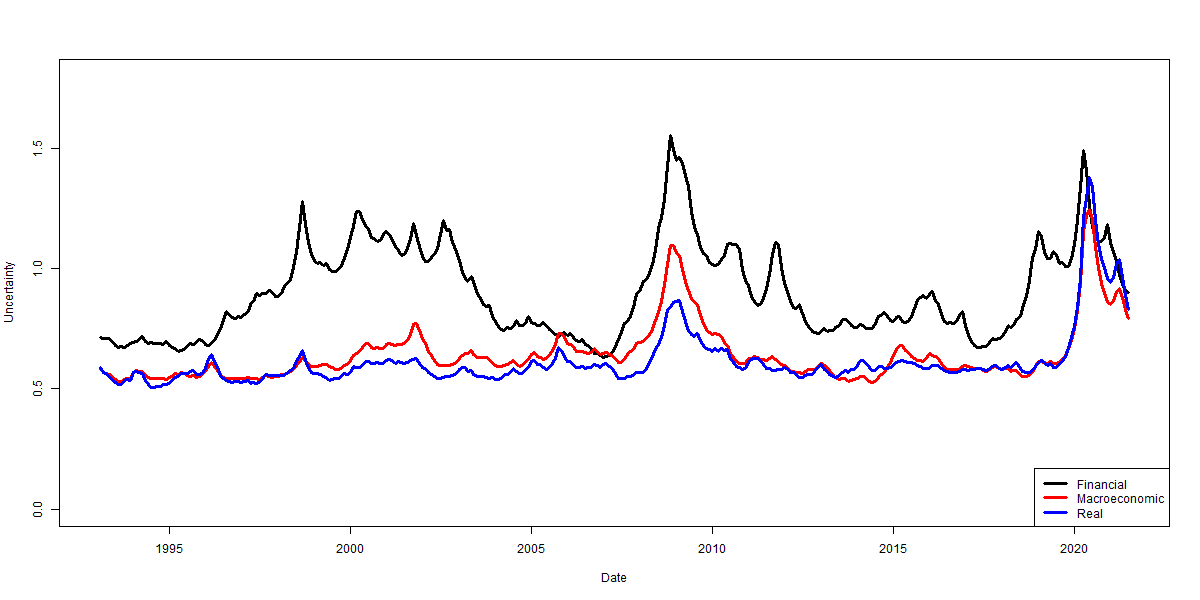}}  \\
             \multicolumn{2}{c}{(a) 1-month ahead} \\
             \includegraphics[width=8cm]{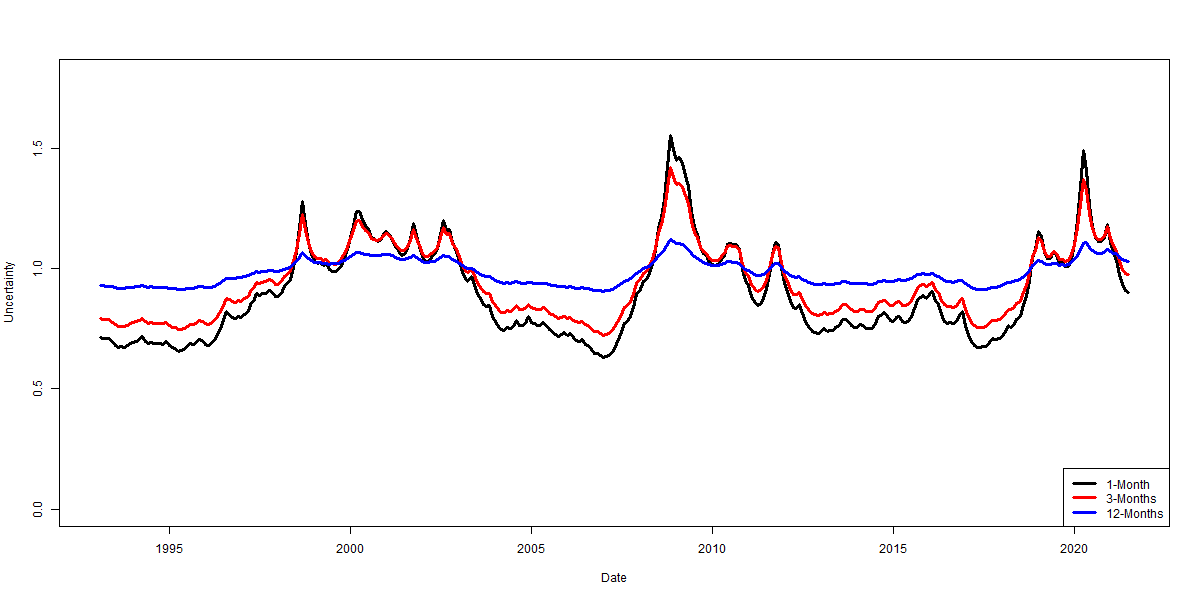} &
             \includegraphics[width=8cm]{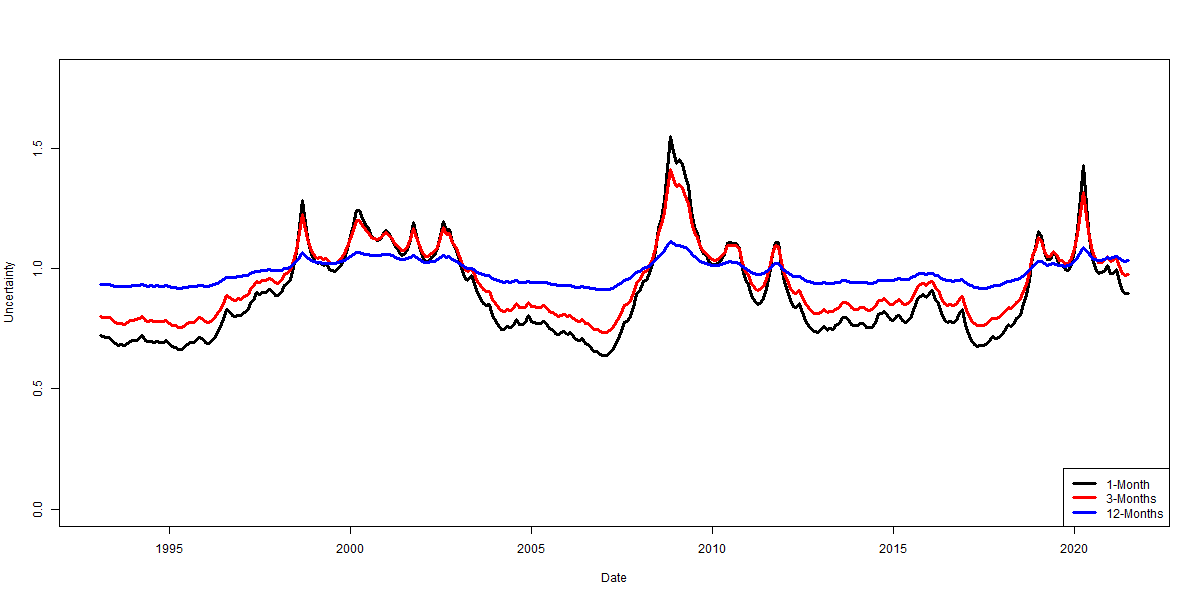} \\
             (b) Financial Uncertainty (FIN) & (c) Financial Uncertainty (Economic) \\
             \includegraphics[width=8cm]{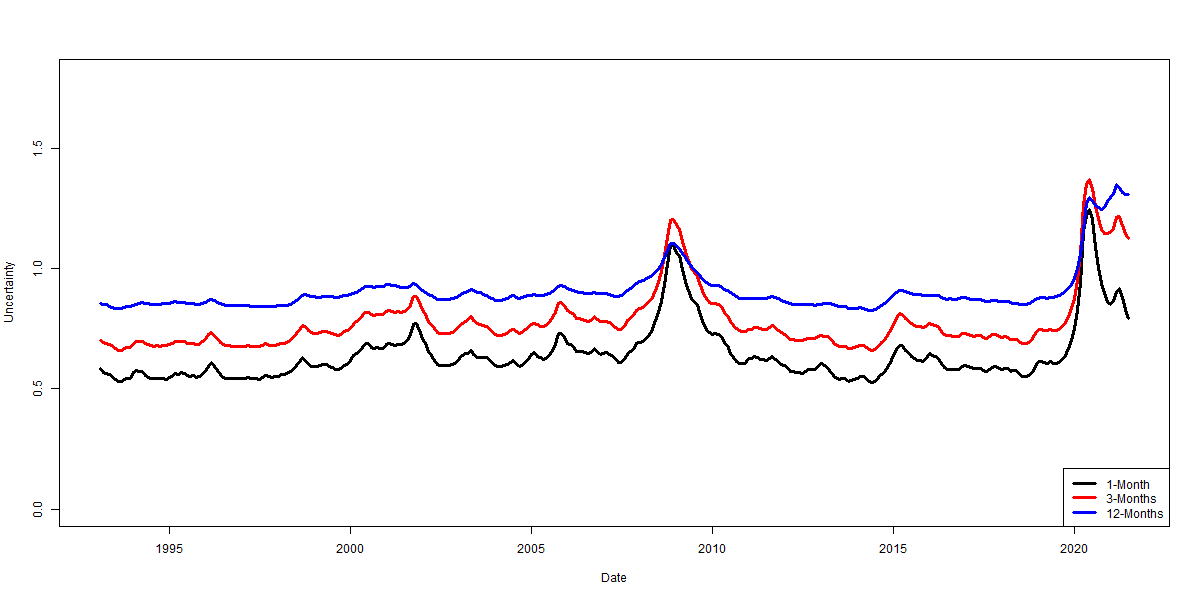} &
             \includegraphics[width=8cm]{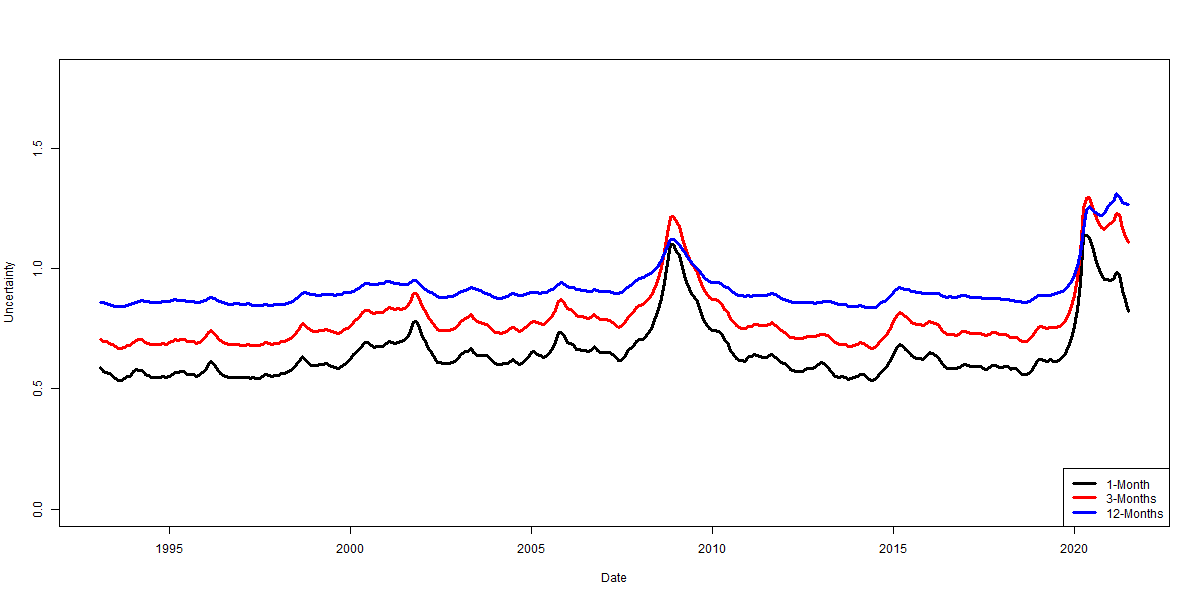} \\
             (d) Macroecnomic Uncertainty (MAC) & (e) Macroeconomic Uncertainty (Economic) \\
             \includegraphics[width=8cm]{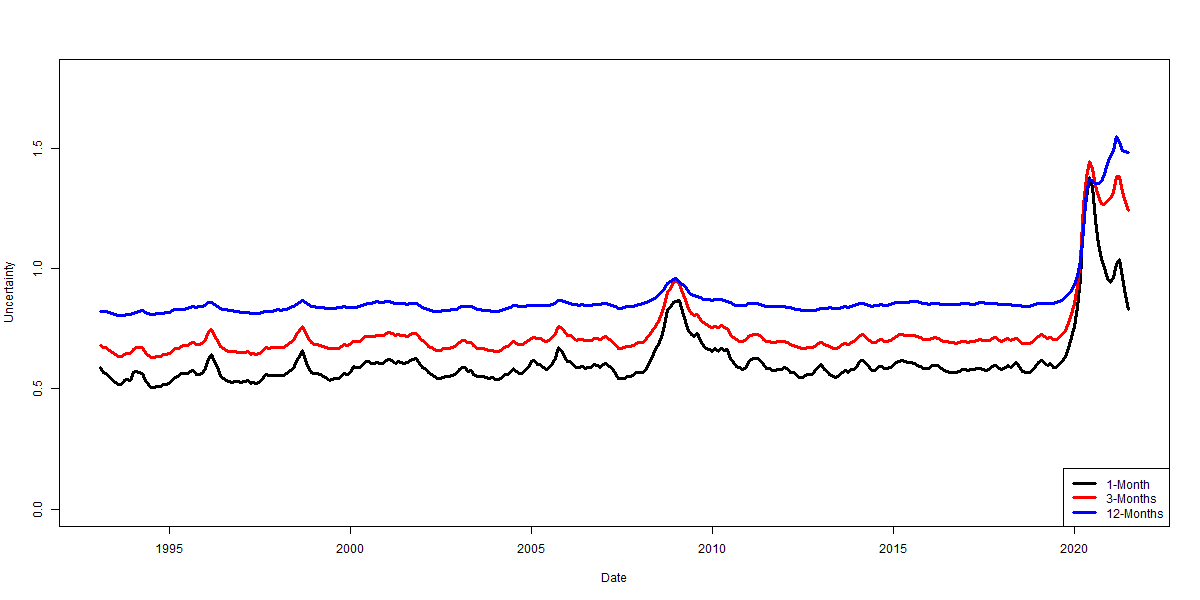} &
             \includegraphics[width=8cm]{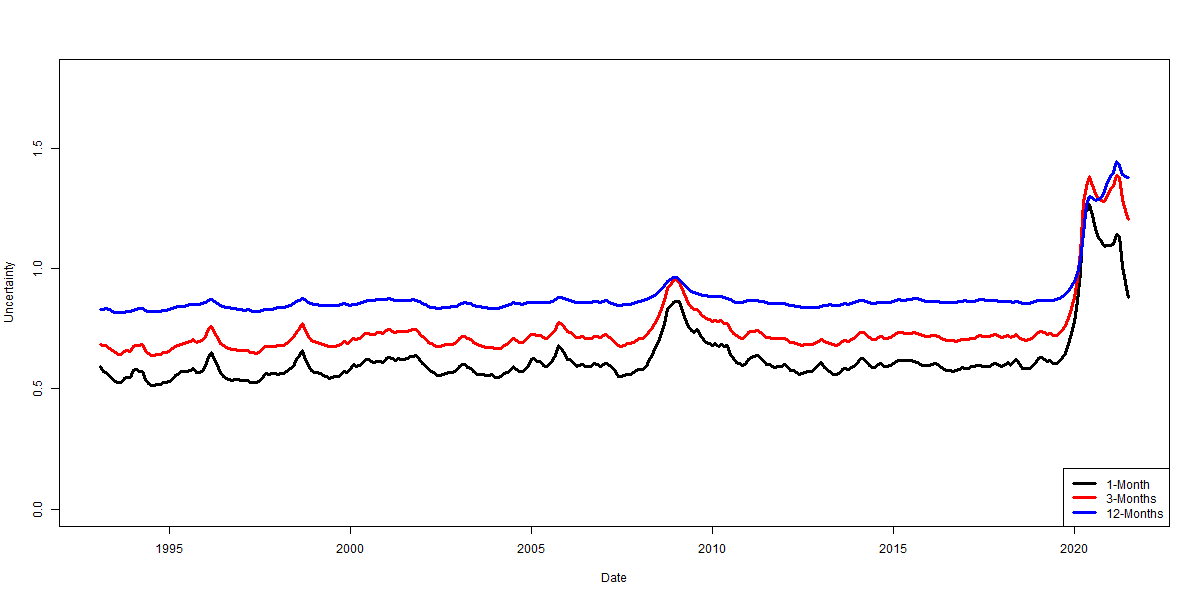} \\
             (f) Real Uncertainty (REA) & (g) Real Uncertainty (Economic) \\
        \end{tabular}
    \end{center}
\footnotesize{Notes: Figures plot uncertainty indexes of \cite{jurado2015measuring} as obtained from the website of Sydney Ludvigson \texttt{https://www.sydneyludvigson.com/macro-and-financial-uncertainty-indexes}. Uncertainty indexes based upon the financial (FIN), macroeconomic (MAC) and real (REA) uncertainty are shown at the 1, 3 and 12 month horizon. Panel (a) contrasts the one month ahead uncertainty index as calculated from the three sets of time series. Panels (b), (d) and (f) use all available time series, whilst (c), (e) and (g) use only those classed as ``economic''. Full details of construction are provided in \cite{jurado2015measuring}. Sample period is based upon the availability of data on the Russell 2000 stock index and runs from 1st January 1993 to 30th June 2021. All data is monthly.}
\end{figure}

Figure \ref{fig:aunc} presents comparisons between the three one-month ahead uncertainty measures in panel (a), and then the three different time spans for each uncertainty measure in panels (b) to (g). Immediately it may be seen that the financial uncertainty measure remains the most volatile, as observed in \cite{jurado2015measuring}. We see that there are three spikes in the FIN series, 2 in the MAC and just one in the REA. In every case the one-month ahead is the most volatile, the 12-month ahead measure being far more persistent through the time periods.

However, in the Covid-19 period we see that all three uncertainty measures have very similar behaviour\footnote{A further useful commentary on recent uncertainty is provided in \cite{altig2020economic}, with a discussion on the differentials between the \cite{jurado2015measuring} and other approaches to capturing uncertainty in the light of Covid-19.}. To restrict the effect of health related variables, the ``Economic'' series were created. Contrasting the right hand plot on each row with the total uncertainty we see that there is very little difference in the series. Following the literature we continue therefore to use total uncertainty in the main paper. Results using the economic measure are qualitatively similar and are available on request.

\begin{sidewaystable}
    
    \begin{center}
        \caption{Correlation of Uncertainty Measures}
        \label{tab:aucor}
        \begin{footnotesize}
        \begin{tabular}{l c c c c c c c c c c c c c c c c c c}
        \hline
        & \multicolumn{6}{c}{Financial Uncertainty} & \multicolumn{6}{c}{Macroeconomic Uncertainty} & \multicolumn{6}{c}{Real Uncertainty}\\
        & \multicolumn{3}{c}{Total} & \multicolumn{3}{c}{Economic} & \multicolumn{3}{c}{Total} & \multicolumn{3}{c}{Economic} & \multicolumn{3}{c}{Total} & \multicolumn{3}{c}{Economic} \\
        & 1 & 3 & 12 & 1 & 3 & 12& 1 & 3 & 12& 1 & 3 & 12& 1 & 3 & 12& 1 & 3 & 12\\
        \hline
       FIN1 & 1&0.999&0.988&0.994&0.995&0.987&0.553&0.559&0.548&0.555&0.561&0.55&0.437&0.446&0.420&0.465&0.467&0.43\\
FIN3&0.999&1&0.992&0.992&0.994&0.991&0.554&0.561&0.551&0.556&0.563&0.553&0.435&0.445&0.419&0.463&0.465&0.428\\
FIN12&0.980&0.989&1&0.976&0.982&0.998&0.559&0.567&0.564&0.560&0.567&0.565&0.425&0.437&0.413&0.451&0.455&0.421\\
EFIN1&0.993&0.990&0.965&1&0.999&0.982&0.528&0.534&0.523&0.530&0.536&0.525&0.408&0.418&0.390&0.437&0.439&0.400\\
EFIN3&0.994&0.994&0.977&0.998&1&0.987&0.532&0.538&0.528&0.534&0.54&0.530&0.410&0.420&0.392&0.438&0.441&0.402\\
EFIN12&0.98&0.988&0.996&0.975&0.986&1&0.546&0.554&0.551&0.547&0.554&0.552&0.410&0.422&0.396&0.435&0.439&0.405\\
MAC1&0.666&0.662&0.658&0.608&0.609&0.615&1&0.996&0.967&0.999&0.996&0.970&0.758&0.764&0.734&0.754&0.765&0.742\\
MAC3&0.628&0.632&0.651&0.562&0.570&0.602&0.978&1&0.982&0.996&0.999&0.984&0.741&0.758&0.741&0.738&0.759&0.748\\
MAC12&0.517&0.531&0.585&0.438&0.459&0.529&0.872&0.954&1&0.966&0.981&0.999&0.688&0.720&0.732&0.683&0.719&0.734\\
EMAC1&0.671&0.669&0.671&0.613&0.616&0.628&0.992&0.986&0.898&1&0.996&0.969&0.758&0.765&0.735&0.758&0.768&0.744\\
EMAC3&0.639&0.642&0.660&0.574&0.583&0.613&0.974&0.998&0.950&0.99&1&0.983&0.744&0.761&0.744&0.743&0.764&0.752\\
EMAC12&0.549&0.561&0.609&0.473&0.492&0.556&0.895&0.968&0.998&0.921&0.967&1&0.689&0.719&0.729&0.686&0.720&0.733\\
REA1&0.520&0.524&0.545&0.433&0.443&0.480&0.897&0.924&0.909&0.887&0.906&0.905&1&0.985&0.913&0.995&0.982&0.922\\
REA3&0.446&0.458&0.509&0.354&0.374&0.442&0.819&0.902&0.961&0.834&0.889&0.947&0.961&1&0.959&0.982&0.995&0.964\\
REA12&0.329&0.349&0.424&0.237&0.265&0.359&0.686&0.812&0.939&0.717&0.802&0.916&0.865&0.969&1&0.911&0.954&0.997\\
EREA1&0.518&0.524&0.554&0.429&0.442&0.488&0.877&0.927&0.937&0.886&0.917&0.933&0.988&0.981&0.908&1&0.986&0.925\\
EREA3&0.461&0.473&0.521&0.370&0.389&0.455&0.823&0.906&0.962&0.843&0.896&0.951&0.959&0.998&0.966&0.985&1&0.965\\
EREA12&0.350&0.370&0.441&0.259&0.286&0.376&0.707&0.829&0.947&0.738&0.819&0.925&0.880&0.976&0.999&0.922&0.974&1\\
\hline 
        \end{tabular}
        \end{footnotesize}
    \end{center}
\footnotesize{Notes: Figures report correlations between the stated uncertainty indexes of \cite{jurado2015measuring}. Figures below the diagonal are Pearson correlation coefficients, whilst those above the diagonal are Spearman's rank correlation coefficients. The indexes codes are formed of three letters to indicate the family, being either financial uncertainty (FIN), macroeconomic uncertainty (MAC), or real uncertainty (REA). Each code includes a number to indicate the number of months ahead, being either 1, 3 or 12. Finally for those using the ``economic'' series a prefix ``E'' is added. Sample from 1st January 1993 to 28th June 2021.}
\end{sidewaystable}

Formalising the message from Figure \ref{fig:aunc}, Table \ref{tab:aucor} presents a full 18x18 correlation matrix for the uncertainty indexes. We denote the economic series with an E prefix. We see a very clear blocking of both the Pearson correlations in the lower left triangle, and the Spearman's ranked correlation ceofficients above the diagonal. Of the three sets of measure the within variation for MAC is the highest. Correlation between $MAC1$ and $FIN1$ is 0.666, whilst the rank correlation of the series is little over 0.5. This difference explains the potential for the persistence norms to provide additional information to financial uncertainty without necessarily giving extra explanatory power in regressions with $MAC1$ as the dependent variable. This is what we find in the main paper.

The brief exposition here is consistent with the discussion in \cite{jurado2015measuring} and the interested reader is referred to the original paper for further discussion of the relationships between the uncertainty indexes. Whilst Covid-19 has seen different behaviour from REA, MAC1 shows only a short spike and FIN is already looking to be beyond its spike at all horizons. There is undoubtedly an interesting research agenda to understand the rationales for the observed responses. However, such an exploration of Covid-19 responses is beyond the focus of this paper.

\section{Correlation Between Clouds}

In the regressions we treat each of our four cloud lengths independently. This follows since interpretation requires that the volatility and norms be computed on the same data. However, the results are naturally compared across the 1-month, 3-month, 6-month and 12-month clouds. In this short appendix we therefore present the correlation between the three cloud based measures in Table \ref{tab:corcloud}

\begin{table}
    \begin{center}
        \caption{Correlations between cloud based measures}
        \label{tab:corcloud}
        \begin{tabular}{l c c c c c c c c c c c c}
        \hline
        Time & \multicolumn{4}{c}{Volatility $\bar{\sigma}$} & \multicolumn{4}{c}{Volatility $L_{11}$} & \multicolumn{4}{c}{Volatility $L_{12}$} \\
        & 1m & 3m & 6m & 12m & 1m & 3m & 6m & 12m & 1m & 3m & 6m & 12m\\
        \hline
         1m & 1&0.857&0.779&0.715&1&0.318&0.245&0.188&1&0.281&0.237&0.169\\
         3m & 0.858&1&0.92&0.849&0.322&1&0.777&0.661&0.362&1&0.771&0.648\\
         6m & 0.723&0.893&1&0.931&0.283&0.831&1&0.835&0.301&0.835&1&0.830\\
        12m & 0.601&0.735&0.867&1&0.272&0.733&0.879&1&0.254&0.709&0.859&1\\
        \hline
        \end{tabular}
    \end{center}
    \footnotesize{Notes: Correlations of the stated cloud measure between observations from the four cloud lengths. 1m, 3m, 6m and 12m refer to the number of months of data in the point cloud over which the persistence norms and volatility are calculated. Clouds are formed from the daily log returns on the S\&P 500, Dow Jones Industrial Average, NASDAQ and Russell 2000 stock market indexes. Figures below the diagonal are the Pearson correlation coefficient and figures above the diagonal are the Spearman's rank correlation coefficient. Overall sample from 1st January 1993 to 28th June 2021.}
\end{table}

Within the three measures taken from our clouds we can see that the between lengths correlations are much higher for volatility than they are for the persistence norms. The correlation between the 1-month cloud volatility and the 12-month cloud volatility is 0.601, but the corresponding figure for the $L_{11}$ norms is 0.272 and for the $L_{12}$ norms is 0.254. Between the 1-month and 3-month clouds the correlations are 0.858 for the volatility, but just 0.322 for $L_{11}$ and 0.362 for $L_{12}$. When we see the differences in the regression results it is understandable because of these low correlations. From the construction of persistence norms described in Section 2 of the main paper, it follows that more points produces a denser cloud and will lead to features being born, and dying, at lower filtration radii than is the case for the sparse 1-month cloud.

\section{Four Market Cloud with $L2$ Norms}

In the main paper we follow \cite{gidea2018topological} to use the $L_1$ norms from the point cloud of index returns. Table 1 shows that the correlation between the $L_1$ and $L_2$ norms is high and so we may expect that the regression results are also similar. Through this appendix we demonstrate that there is consistency to the inference if using the $L_2$ norms.

\begin{figure}
    \begin{center}
        \caption{Persistence norms, volatility and uncertainty time series plots}
        \label{fig:unc2}
        \begin{tabular}{c c}
            \includegraphics[width=8cm]{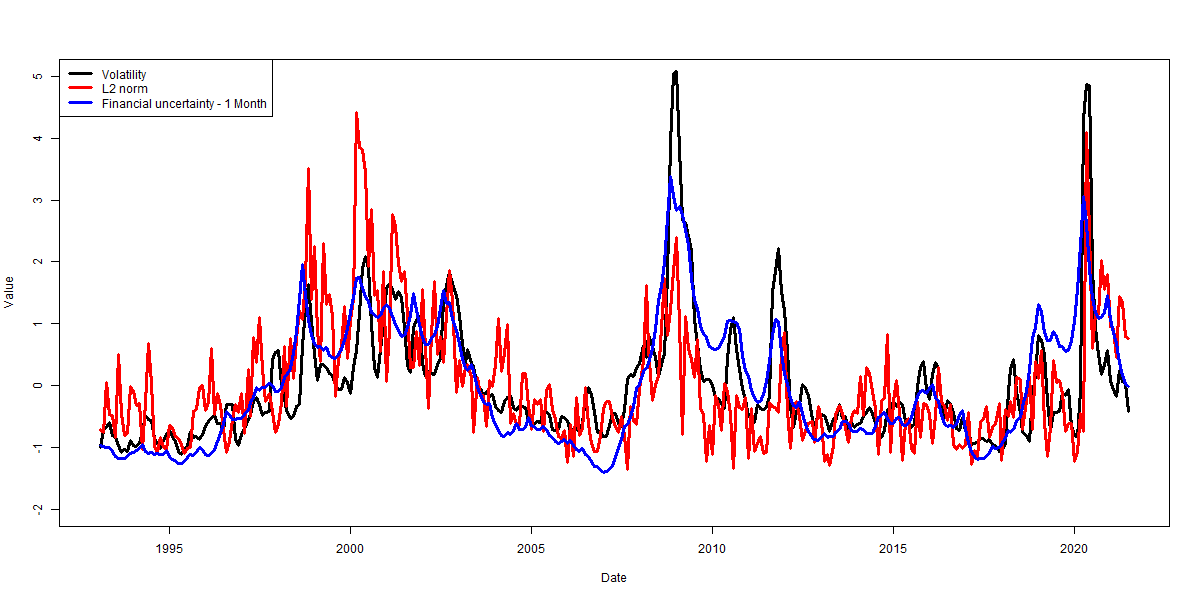} & \includegraphics[width=8cm]{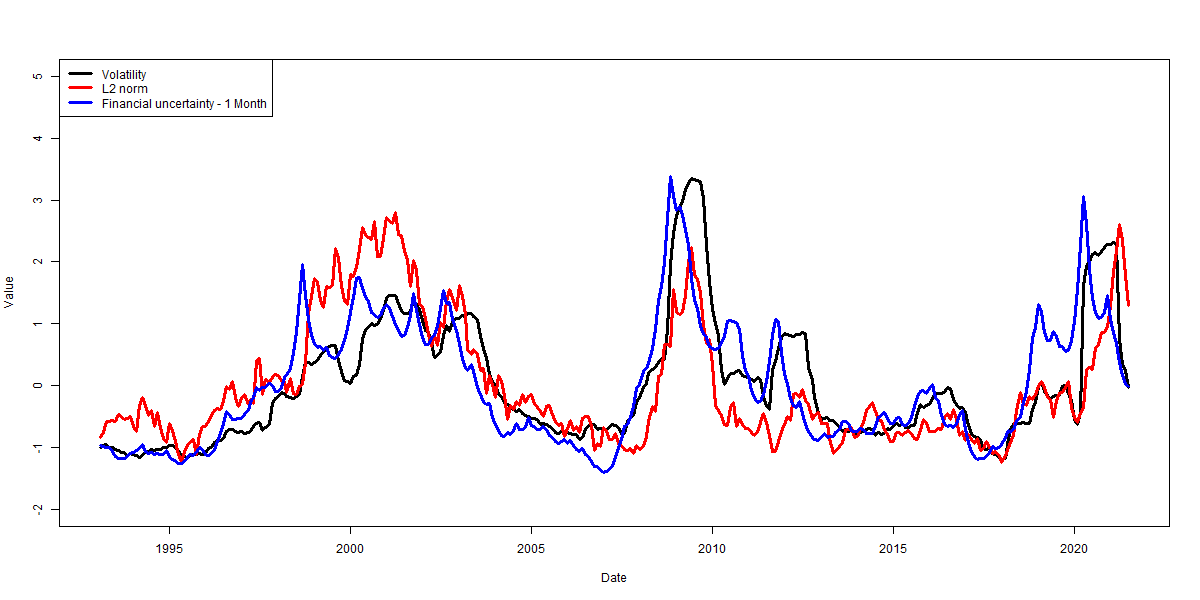}  \\
            (a) $FIN1$  3 months $L_2$ & (b) $FIN1$ 12 months $L_2$ \\ 
            \includegraphics[width=8cm]{dtudafinunc1l11.png} & \includegraphics[width=8cm]{dtufafinunc1l11.png}  \\
            (c) $FIN1$  3 months  $L_1$& (d) $FIN1$ 12 months  $L_1$\\ 
            \includegraphics[width=8cm]{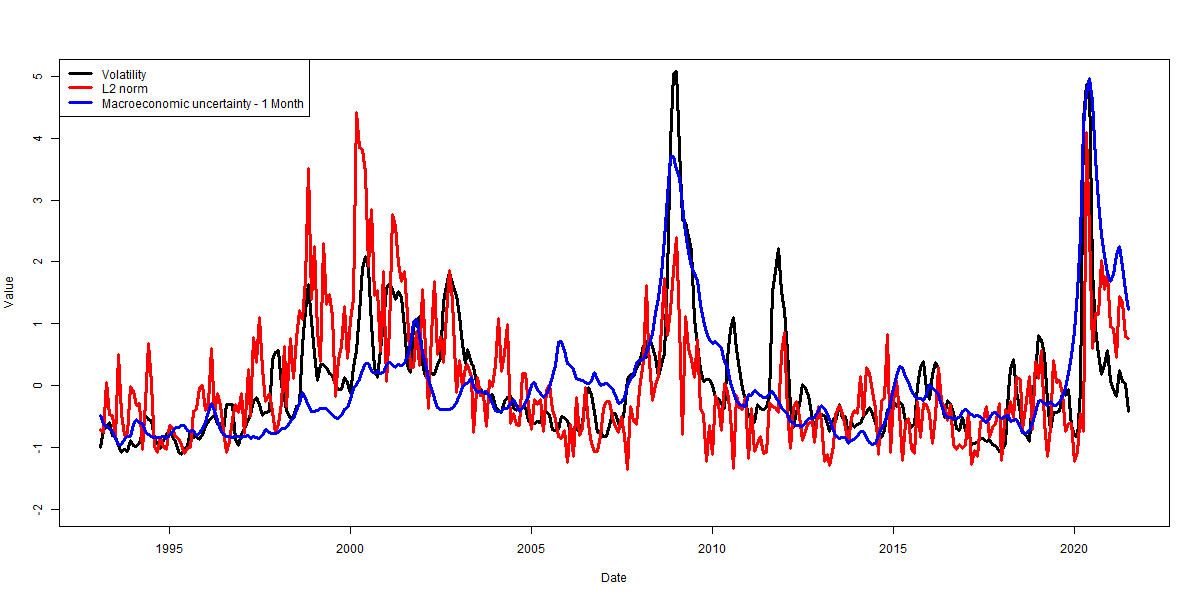} & \includegraphics[width=8cm]{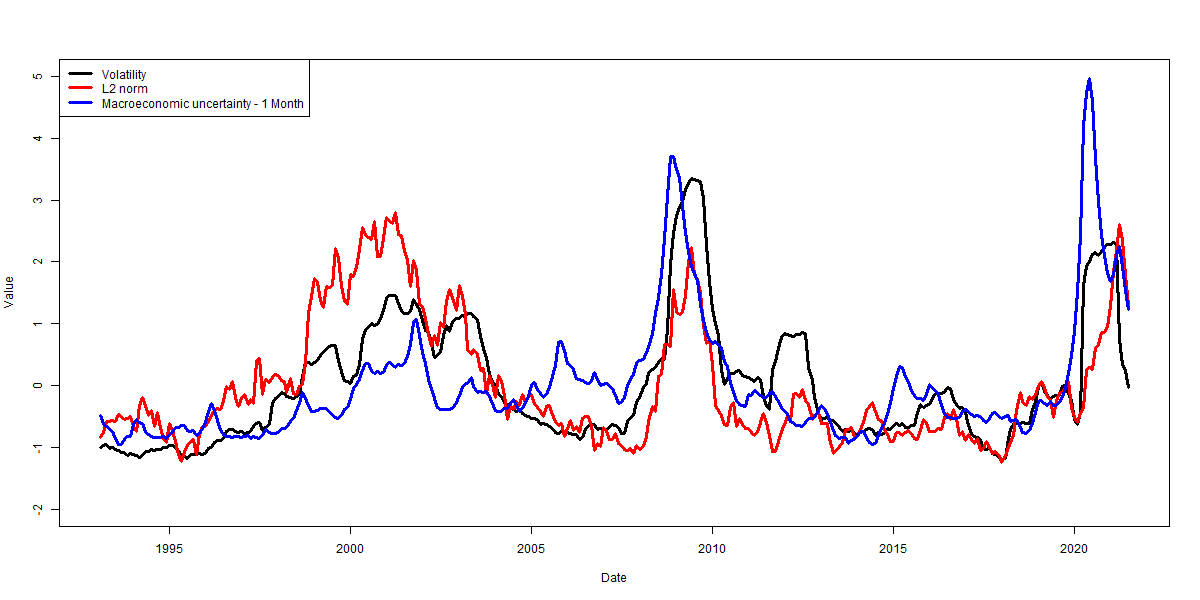}  \\
            (e) $MAC1$ 3 months $L_2$ & (f) $MAC1$ 12 months $L_2$  \\
            \includegraphics[width=8cm]{dtudamacunc1l11.png} & \includegraphics[width=8cm]{dtufamacunc1l11.png}  \\
            (g) $MAC1$ 3 months $L_1$ & (h) $MAC1$ 12 months $L_1$ \\
        \end{tabular}
    \end{center}
\footnotesize{Notes: Figures plot the financial (FIN) and macroeconomic (MAC) uncertainty index of \cite{jurado2015measuring}. Each plot uses the 1 month ahead uncertainty index $FIN1$ and $MAC1$. These indices are plotted as blue lines. On each plot the average return volatility of the four indexes within the cloud; being the S\&P500, Dow Jones Industrial Average, NASDAQ and the Russell 2000 index. Average volatility is plotted in black. The dimension one persistence norms from these clouds are shown in red. Panels (a), (b), (e) and (f) show $L_2$ norms, whilst panels (c), (d), (g) and (h) present $L_1$ Values are standardised by subtracting the mean value and dividing by the standard deviation. Size of window used in the estimation is provided below each plot. Sample from 1st January 1993 to 28th June 2021.}
\end{figure}

Figure \ref{fig:unc2} can be directly compared with Figure 1 of the main paper. To highlight the similarity below each plot of the $L_2$ norm we place the corresponding $L_1$ plot. Thus panels (a), (b), (e) and (f) are the $L_2$ plots and (c), (d), (g) and (h) replicate those from the main paper in showing $L_1$ norms. The main contrasts appear in the ``dot-com'' bubble period towards the left of the plot. Panels (e) and (g) show this well in the tall spike in the $L_2$ contrasting with a shorter spike in the $L_1$. In Panels (b) and (d) we can see a small first spike in the early part of the global financial crisis for the $L_2$ norm, but this is absent in the $L_1$. Both the $L_1$ and $L_2$ norms then have a taller spike at the same time that the volatility reaches its maximum. Through the rest of the panels behaviour is as similar, as the high correlation between the series would suggest. Exploring the rationale for these differences would be an interesting task for further research. 

\subsection{Persistence Norms}

Regression of persistence norms on the uncertainty indices of \cite{jurado2015measuring} and the average volatility index returns tests whether these financial market variables are able to span the persistence norms. Formally we estimate a total of 288 models which are based upon three base regressions:
\begin{align}
    \mathtt{Model 12:} & L_{12} = \alpha_1 + \beta_{11}FIN1 + \epsilon \\
    \mathtt{Model 22:} & L_{12} = \alpha_2 + \beta_{21}\bar{\sigma} + \epsilon \\
    \mathtt{Model 32:} & L_{12} = \alpha_3 + \beta_{31}FIN1 + \beta_{32}\bar{\sigma} + \epsilon 
\end{align}
Where $L_{12}$ is the dimension one $L_2$ persistence norm, $\sigma$ is the standard deviation of S\&P 500 returns and $FIN1$ is the one month ahead financial uncertainty index of \cite{jurado2015measuring}. In all cases $\epsilon$ is the error term. In all estimations \cite{newey1987simple} adjusted standard errors are employed to account for autocorrelation and heteroskedasticity inherent in financial time series. This set up is identical to that in the main paper, but here we are reporting $L_2$ norms instead of $L_1$. The addition of the 2 to the model allows distinguishing with models 1 and 2 in the main paper. We do not add any additional subscripts to the coefficients for ease of reading.

As in the main paper Models 12, 22 and 32 are just the building blocks for a series of estimations that must be performed. Again, for each measure of uncertainty there are 4 point clouds to consider. Three uncertainty indices at three different lags, and with both a standard and economic version for each of the 9, this all means 72 combinations arise from Model 12. Across the three models the total number of combinations is thus 216. This exposition focuses on FIN and MAC at the 1 month ahead horizon, but as noted results for other uncertainty indexes and time horizons are available on request from the authors.

\begin{sidewaystable}
    \begin{center}
        \caption{Determinants of $L_2$ persistence norms}
        \label{tab:reg12}
        \begin{tabular}{ll l c c c c c c c c c c c c c}
        \hline
             &&& \multicolumn{4}{l}{Model 12} & \multicolumn{4}{l}{Model 22}& \multicolumn{4}{l}{Model 32}\\
             &&& 1m & 3m & 6m & 12m & 1m & 3m & 6m & 12m & 1m & 3m & 6m & 12m\\
             \hline
             \multicolumn{12}{l}{Panel (a): Financial Uncertainty ($FIN1$): }\\
             & Constant & $\alpha$ & 0.012&-0.162&-0.189&-0.109&0.092&0.154&0.221&0.292&0.000&-0.062&-0.051&0.085\\
			&&&(0.303)&(2.112)&(1.531)&(0.428)&(7.129)&(5.528)&(4.491)&(3.347)&(0.008)&(0.620)&(0.377)&(0.499)\\
			&Uncertainty & $FIN1$&0.113&0.569&0.790&0.921&&&&&0.139&0.350&0.465&0.369\\
			&&&(2.396)&(5.799)&(4.861)&(2.561)&&&&&(2.543)&(2.372)&(2.135)&(1.340)\\
			& Volatility & $\bar{\sigma}$&&&&&0.020&0.173&0.262&0.365&-0.010&0.085&0.134&0.258\\
			&&&&&&&(1.538)&(5.576)&(4.964)&(3.566)&(-0.631)&(2.467)&(2.460)&(4.058)\\
			&\multicolumn{2}{l}{Adjusted $R^2$}&0.034&0.362&0.440&0.442&0.013&0.343&0.426&0.511&0.036&0.392&0.477&0.538\\
            &&&&&&\\
             \multicolumn{9}{l}{Panel (b): Macroeconomic Uncertainty ($MAC1$): }\\
             & Constant & $\alpha$ & 0.041&-0.017&0.031&0.179&0.092&0.154&0.221&0.292&0.052&0.183&0.347&0.521\\
			&&&(0.747)&(0.314)&(0.294)&(0.439)&(7.129)&(5.528)&(4.491)&(3.347)&(0.802)&(1.484)&(1.770)&(2.854)\\
			&Uncertainty & $MAC1$&0.114&0.570&0.764&0.841&&&&&0.075&-0.059&-0.269&-0.504\\
			&&&(1.286)&(7.047)&(4.728)&(1.227)&&&&&(0.604)&(0.226)&(0.547)&(1.100)\\
			&Volatility & $\bar{\sigma}$&&&&&0.020&0.173&0.262&0.365&0.012&0.181&0.303&0.446\\
			&&&&&&&(1.538)&(5.576)&(4.964)&(3.566)&(0.631)&(3.400)&(2.458)&(2.854)\\
			&\multicolumn{2}{l}{Adjusted $R^2$}&0.014&0.144&0.162&0.145&0.013&0.343&0.426&0.511&0.017&0.344&0.436&0.538\\
             \hline
        \end{tabular}
    \end{center}
\footnotesize{Notes: Regressions of the stated outcome with the other two variables as explanatory factors. 3m and 6m refers to the number of months of data in the point cloud over which the persistence norms and volatility are calculated. Point clouds are the daily returns on the S\&P 500, Dow Jones Industrial Average, NASDAQ and Russell 2000 indexes. Uncertainty indexes are the 1 month ahead estimates from \cite{jurado2015measuring} as downloaded from the website of Sydney Ludvigson. Figures are the estimated coefficients and figures in parentheses are the associated \cite{newey1987simple} adjusted t-statistics. Model 12 has $L_{12} = \alpha_1 + \beta_{12}FIN1 + \epsilon$. Model 22 has $L_{12} = \alpha_2 + \beta_{21}\bar{\sigma} + \epsilon$. Model 32 has $L_{2} = \alpha_3 + \beta_{31}FIN1 + \beta_{32}\bar{\sigma} + \epsilon$. Models 12, 22 and 32 represent the base specification upon which all regression models are built. Specifications show $FIN1$ as the one month ahead financial uncertainty (FIN) index. Where macroeconomic uncertainty (MAC) is used then the equations for the four models are duly updated to replace $FIN1$ with $MAC1$. In each case $\bar{\sigma}$ is the average standard deviation of the four index returns in the window and $\epsilon$ is the error term. Estimation employs \cite{newey1987simple} robust standard errors to control for autocorrelation and heteroskedasticity inherent in these time series. Sample from 1st January 1993 to 28th June 2021.}
\end{sidewaystable}

Table \ref{tab:reg12} reports estimates from the regressions with \cite{newey1987simple} robust standard errors. Panel (a) shows that $FIN1$ is significant in both models 12 and 32. Volatility is significant in model 22 and here in model 32. Herein we see the main difference between $L_1$ and $L_2$ comes in the insignificance of $FIN1$ when volatility is included. Caution is urged because of the high correlation between volatility and $FIN1$, but it may also be noted that $L_{12}$ has a stronger correlation to $\bar{\sigma}$ than $L_{11}$ in Table 1 of the main paper. Further when using only volatility as the explanatory variable for the $L_{12}$ norm, the constant term is highly significant, but this is not the case when uncertainty appears in the specification. Recalling that the same observation was found for the $L_1$ norm we have the consistency between the results of Table \ref{tab:reg12} and Table 2 of the main paper.

Further consistency is found in the results of the 1-month cloud, with the adjusted r-squared being much lower than the longer clouds. As a month only has 22 trading days on average it follows that the cloud being used is quite small. At a smaller size the impact of an extreme value is much larger, the distance between the extremes and the next nearest point in the cloud will then influence the feature construction in the rips complex. Although the measures from the 1-month cloud are consistent with those from longer clouds on average, we are directed to use longer clouds if we wish to explain the behaviour of the persistence norms.

\subsection{Uncertainty}

Reversing the question we regress the uncertainty indexes on the persistence norms and volatility. 
\begin{align}
    \mathtt{Model 42:} & FIN1 = \alpha_4 + \beta_{41}L_{12} + \epsilon \\
    \mathtt{Model 52:} & FIN1 = \alpha_5 + \beta_{52}\bar{\sigma} + \epsilon\\
    \mathtt{Model 62:} & FIN1 = \alpha_6 + \beta_{61}L_{12} + \beta_{62}\bar{\sigma} + \epsilon
\end{align}
Here again $FIN1$ is the one month ahead financial uncertainty from \cite{jurado2015measuring}, $L_{11}$ is the dimension 1 $L_1$ persistence norm and $\bar{\sigma}$ is the mean standard deviation from the four return series in the cloud. $\epsilon$ is the error term. Models are estimated with \cite{newey1987simple} robust standard errors. Once more we add a 2 to the name of the model to ensure correct referencing against models 4, 5 and 6 of the main paper. It must be noted at this point that results from model 52 will be identical to model 5 in the main paper as uncertainty and volatility are not affected by the choice of persistence norms.

\begin{sidewaystable}
    \begin{center}
        \caption{Explaining the uncertainty index}
        \label{tab:reg42}
        \begin{tabular}{ll l c c c c c c c c c c c c c}
        \hline
             &&& \multicolumn{4}{l}{Model 42} & \multicolumn{4}{l}{Model 52}& \multicolumn{4}{l}{Model 62}\\
             &&& 1m & 3m & 6m & 12m & 1m & 3m & 6m & 12m & 1m & 3m & 6m & 12m\\
             \hline
             \multicolumn{12}{l}{Panel (a): Financial Uncertainty ($FIN1$): }\\
             & Constant & $\alpha$ & 0.867&0.678&0.610&0.556&0.667&0.617&0.584&0.560&0.651&0.584&0.542&0.517\\
			&&&(13.62)&(18.46)&(11.80)&(6.067)&(20.77)&(16.46)&(15.70)&(8.128)&(21.07)&(16.27)&(15.62)&(6.583)\\
			&Norm&$L_{12}$&0.304&0.637&0.557&0.479&&&&&0.168&0.212&0.192&0.149\\
			&&&(3.063)&(8.153)&(5.795)&(3.685)&&&&&(2.474)&(3.395)&(2.467)&(1.252)\\
			&Volatility & $\bar{\sigma}$&&&&&0.213&0.251&0.274&0.290&0.210&0.215&0.224&0.236\\
			&&&&&&&(11.02)&(10.50)&(7.333)&(3.727)&(10.62)&(10.69)&(5.770)&(2.662)\\
			&\multicolumn{2}{l}{Adjusted$R^2$}&0.034&0.362&0.440&0.442&0.552&0.646&0.663&0.622&0.563&0.672&0.693&0.643\\
             &&&&&&&\\
             \multicolumn{9}{l}{Panel (b): Macroeconomic Uncertainty ($MAC1$): }\\
             & Constant & $\alpha$ & 0.631&0.556&0.533&0.520&0.528&0.492&0.468&0.455&0.524&0.496&0.482&0.487\\
			&&&(10.57)&(14.62)&(8.537)&(8.841)&(31.34)&(14.50)&(9.186)&(7.774)&(29.21)&(12.69)&(10.77)&(9.884)\\
			&Norm &$L_{12}$&0.120&0.252&0.212&0.173&&&&&0.053&-0.021&-0.063&-0.109\\
		&&	&(0.634)&(1.245)&(1.041)&(1.383)&&&&&(0.565)&(0.225)&(0.792)&(0.949)\\
			&Volatility & $\bar{\sigma}$ &&&&&0.105&0.134&0.153&0.161&0.104&0.138&0.169&0.201\\
			&&&&&&&(5.057)&(2.614)&(2.311)&(2.268)&(4.958)&(3.191)&(2.495)&(1.944)\\
			&\multicolumn{2}{l}{Adjusted $R^2$}&0.014&0.144&0.162&0.145&0.342&0.467&0.520&0.486&0.345&0.468&0.528&0.514\\
             \hline
        \end{tabular}
    \end{center}
\footnotesize{Notes: Regressions of the stated outcome with the other two variables as explanatory factors. 1m, 3m, 6m and 12m refers to the number of months of data in the point cloud over which the persistence norms and volatility are calculated. Uncertainty indexes are the 1 month ahead estimates from \cite{jurado2015measuring} as downloaded from the website of Sydney Ludvigson. Clouds are formed from the daily log returns on the S\&P 500, Dow Jones Industrial Average, NASDAQ and Russell 2000 stock market indexes. Figures are the estimated coefficients and figures in parentheses are the associated \cite{newey1987simple} adjusted t-statistics. Model 42 has $FIN1 = \alpha_4 + \beta_{41}L_{12} + \epsilon$. Model 52 has $FIN1 = \alpha_5 + \beta_{51}\bar{\sigma} + \epsilon$. Model 62 has $FIN1 = \alpha_6 + \beta_{61}L_{12} + \beta_{62}\bar{\sigma}+\epsilon$. Specifications show $FIN1$ as the one month ahead financial uncertainty (FIN) index. Where macroeconomic uncertainty (MAC) is used then the equations for the four models are duly updated to replace $FIN1$ with $MAC1$. In each case $\bar{\sigma}$ is the average standard deviation of the returns on the four indices in the window. $\epsilon$ is the error term. Estimation employs \cite{newey1987simple} robust standard errors to control for autocorrelation and heteroskedasticity inherent in these time series. Sample from 1st January 1993 to 28th June 2021.}
\end{sidewaystable}

Panel (a) of Table \ref{tab:reg42} shows again that the persistence norm is significant as an explanatory variable for $FIN1$. However, the constant terms in Model 42 are also highly significant and the adjusted r-squared values are around 0.40. Volatility has a higher significance and, when combining volatility and norms in Model 62, it is volatility that has the most significant coefficients. Almost two-thirds of the variation in FIN1 may be explained by volatility and persistence norms, a large increase from the just under half that is explained by volatility alone. Model fits using the $L_2$ norm are slightly better than those obtained the $L_1$ norm in the main paper. Importantly the qualitative messages emerging from panel (a) are very similar. 

Panel (b) of Table \ref{tab:reg42} considers MAC and specifically the one-month ahead macroeconomic uncertainty index ($MAC1$). Here, as in the main paper, we see no significance of the persistence norms from clouds of any-length. Because the volatility is not affected by the choice of norms in the model, results for model 52 match those from model 5 exactly. Drawing parallels with the $L_1$ norms in the main paper, we may again see that the results are qualitatively similar. Hence whilst persistence norms add to the fit of FIN they do not provide additional information in the fit of MAC. 

\subsection{High and Low Uncertainty}

As with the main paper, the strong correlation between volatility and uncertainty means that we do not consider high and low norm periods. Here we ask whether there is a different role for norms and volatility in periods of high, or low, uncertainty. To achieve this we split the sample at the median of the respective uncertainty indexes. As neither the uncertainty index, nor the volatility, are affected by the choice of norm the results from model 52 are again identical to those from model 5 in the main paper.

\begin{sidewaystable}
    \begin{center}
        \caption{High and low uncertainty index periods}
        \label{tab:reg42hl}
        \begin{small}
        \begin{tabular}{ll l c c c c c c c c c c c c c}
        \hline
             &&& \multicolumn{4}{l}{Model 4} & \multicolumn{4}{l}{Model 5}& \multicolumn{4}{l}{Model 6}\\
             &&& 1m & 3m & 6m & 12m & 1m & 3m & 6m & 12m & 1m & 3m & 6m & 12m\\
             \hline
             \multicolumn{12}{l}{Panel (a): High Financial Uncertainty ($FIN1$): }\\
             & Constant & $\alpha$ & 1.048&0.933&0.906&0.907&0.877&0.836&0.821&0.822&0.866&0.815&0.792&0.806\\
             &&&(12.24)&(21.47)&(16.83)&(9.791)&(25.29)&(20.84)&(18.08)&(9.144)&(25.38)&(15.24)&(14.47)&(7.227)\\
             & Norms & $L_{12}$&0.105&0.291&0.238&0.175&&&&&0.087&0.086&0.081&0.038\\
             &&&(0.993)&(4.197)&(2.980)&(1.385)&&&&&(1.246)&(1.05)&(1.121)&(0.352)\\
             & Volatility & $\bar{\sigma}$&&&&&0.128&0.151&0.158&0.157&0.127&0.140&0.142&0.146\\
             &&&&&&&(7.316)&(7.887)&(4.930)&(2.144)&(7.013)&(6.518)&(3.896)&(1.708)\\
             & \multicolumn{2}{l}{Adjusted $R^2$}&0.011&0.176&0.178&0.125&0.493&0.528&0.442&0.313&0.501&0.541&0.458&0.317\\

             &&&&&&&\\
             \multicolumn{12}{l}{Panel (b): Low Financial Uncertainty ($FIN1$): }\\
             & Constant & $\alpha$ & 0.739&0.724&0.663&0.588&0.678&0.625&0.579&0.620&0.678&0.625&0.558&0.568\\
             &&&(17.40)&(14.73)&(14.30)&(7.435)&(28.10)&(18.77)&(25.68)&(18.85)&(28.09)&(19.82)&(20.58)&(8.163)\\
             & Norms & $L_{12}$& 0.026&0.069&0.200&0.275&&&&&-0.001&0.000&0.080&0.127\\
             &&&(0.570)&(0.971)&(1.984)&(2.345)&&&&&(0.024)&(0.006)&(0.945)&(0.856)\\
             & Volatility & $\bar{\sigma}$&&&&&0.084&0.151&0.207&0.148&0.084&0.151&0.193&0.125\\
             &&&&&&&(3.626)&(4.062)&(8.092)&(4.490)&(3.584)&(4.117)&(6.217)&(2.787)\\
             & \multicolumn{2}{l}{Adjusted $R^2$}&0.001&0.012&0.102&0.182&0.142&0.257&0.392&0.348&0.142&0.257&0.406&0.379\\

             &&&&&&&&\\
             \multicolumn{9}{l}{Panel (c): High Macroeconomic Uncertainty ($MAC1$): }\\
             & Constant & $\alpha$ & 0.699&0.627&0.615&0.612&0.598&0.549&0.515&0.507&0.583&0.548&0.531&0.540\\
             &&&(10.34)&(11.54)&(7.829)&(9.758)&(24.77)&(16.95)&(11.38)&(7.552)&(27.65)&(12.97)&(12.99)&(9.157)\\
             & Norms & $L_{12}$&0.166&0.233&0.176&0.133&&&&&0.145&0.005&-0.071&-0.121\\
             &&&(1.096)&(0.828)&(0.694)&(0.993)&&&&&(1.762)&(0.037)&(0.757)&(0.940)\\
             & Volatility & $\bar{\sigma}$&&&&&0.089&0.122&0.146&0.149&0.088&0.122&0.164&0.194\\
             &&&&&&&(4.158)&(3.217)&(2.921)&(2.042)&(3.914)&(3.417)&(3.223)&(1.754)\\
             & \multicolumn{2}{l}{Adjusted $R^2$}&0.027&0.133&0.123&0.091&0.281&0.428&0.504&0.407&0.302&0.428&0.516&0.446\\

             &&&&&&&\\
             \multicolumn{9}{l}{Panel (d): Low Macroeconomic Uncertainty ($MAC1$): }\\
             & Constant & $\alpha$ & 0.568&0.556&0.552&0.551&0.553&0.547&0.543&0.537&0.553&0.546&0.543&0.541\\
             &&&(55.27)&(45.45)&(29.97)&(28.49)&(57.69)&(40.03)&(40.67)&(30.85)&(57.13)&(38.22)&(35.09)&(32.75)\\
             & Norms & $L_{12}$&0.030&0.050&0.042&0.032&&&&&0.002&0.009&0.001&-0.015\\
             &&&(1.334)&(2.689)&(1.905)&(1.808)&&&&&(0.096)&(0.422)&(0.041)&(-0.409)\\
             & Volatility & $\bar{\sigma}$&&&&&0.022&0.028&0.031&0.037&0.021&0.026&0.031&0.043\\
             &&&&&&&(3.787)&(3.249)&(3.57)&(2.116)&(3.499)&(2.568)&(2.842)&(1.257)\\
             & \multicolumn{2}{l}{Adjusted $R^2$}&0.018&0.092&0.092&0.072&0.135&0.191&0.222&0.260&0.135&0.192&0.222&0.269\\
             \hline
        \end{tabular}
        \end{small}
    \end{center}
\footnotesize{Notes: Regressions of the stated outcome with the other two variables as explanatory factors. 1m, 3m, 6m and 12m refers to the number of months of data in the point cloud over which the persistence norms and volatility are calculated. Panels (a) and (c) cover periods of high financial and macroeconomic uncertainty respectively. Likewise panels (b) and (d) cover periods where uncertainty is below the sample median. Uncertainty indexes are the 1 month ahead estimates from \cite{jurado2015measuring} as downloaded from the website of Sydney Ludvigson. Clouds are formed from the daily log returns on the S\&P 500, Dow Jones Industrial Average, NASDAQ and Russell 2000 stock market indexes. Figures are the estimated coefficients and figures in parentheses are the associated \cite{newey1987simple} adjusted t-statistics. Model 42 has $FIN1 = \alpha_4 + \beta_{41}L_{12} + \epsilon$. Model 52 has $FIN1 = \alpha_5 + \beta_{51}\bar{\sigma} + \epsilon$. Model 62 has $FIN1 = \alpha_6 + \beta_{61}L_{12} + \beta_{62}\bar{\sigma}+\epsilon$. Specifications show $FIN1$ as the one month ahead financial uncertainty (FIN) index. Where macroeconomic uncertainty (MAC) is used then the equations for the four models are duly updated to replace $FIN1$ with $MAC1$. In each case $\bar{\sigma}$ is the average standard deviation of the returns on the four indices in the window. $\epsilon$ is the error term. Estimation employs \cite{newey1987simple} robust standard errors to control for autocorrelation and heteroskedasticity inherent in these time series. The number of observations in each high and low period is 171. Overall sample from 1st January 1993 to 28th June 2021.}
\end{sidewaystable}

Panel (a) of Table \ref{tab:reg42hl} shows again that persistence norms from the 3-month and 6-month clouds are significant in explaining high periods of FIN. Compared to the results in the main paper the adjusted r-squared is higher, but the inference on the role of norms is qualitatively similar. We also see strong similarity in the significance of the persistence norms after the inclusion of volatility in the model. Here the change to the r-squared value is negligible. It remains the case that the constant terms are highly significant in all models. Panel (b) shows that $L_2$ norms give significance in the 6-month and 12-month clouds but that this does not survive the introduction of volatility into the model. Recognising the links between volatility and norms, attributes this result to volatility. Primarily the inference on FIN is that the results are qualitatively similar.

Turning to $MAC1$ in panel (c) we again see that persistence norms from the financial markets do not explain contemporaneous uncertainty. The only coefficient identified with statistical significance for the $L_2$ norm is in the 1-month point cloud and this is a very volatile signal. With typically 22 points in the cloud caution should be extended about interpreting the significance of the $L_2$ norm in model 62 versus the insignificance of the $L_1$ in model 6 of the main paper. Panel (d) shows again that in low periods of $MAC1$ the persistence norm does pick up a significant coefficient in longer length clouds. However, this significance in model 42 disappears when moving to model 62. Again it is volatility that is the main explanatory variable for $MAC1$. 

\subsection{Summary}

This appendix has repeated the analysis from the main paper using the $L_2$ persistence norm in place of the $L_1$. We have shown that the results are qualitatively similar as would have been expected from the strong correlations in Table 1 of the main paper. Some small differences in the series were observed around the ``dot-com'' bubble and understanding the reasons for these may help evaluate the best choice between $L_1$ and $L_2$ norms in practical applications. It remains the case that the use of $L_1$ after \cite{gidea2018topological} is sufficient to provide inference on the links between persistence norms, volatility and the uncertainty indices of \cite{jurado2015measuring}.

\end{document}